\documentclass[entropy,article,submit,moreauthors,pdftex,10pt,a4paper]{mdpi} 

\firstpage{1} 
\articlenumber{1}
\pubvolume{xx}
\pubyear{2018}
\copyrightyear{2018}
\history{Received: date; Accepted: date; Published: date}
\issuenum{1}
 
\usepackage{dcolumn}
\usepackage{wrapfig}
 \usepackage{mathtools}
  \usepackage{xcolor}
 \usepackage{txfonts}
 \newcommand{\II}{\mathcal{I}}

\mathtoolsset{centercolon}

\newtheorem{statement}[theorem]{Statement}

\newcommand{\ftrunc}[1]{f_{({#1})}}
\newcommand{\ftruncE}[1]{\ftrunc{N}^\equilibrium}

\newcommand{\R}{\varmathbb{R}}
\newcommand{\inta}{\int_{\R^3} \int_0^\infty }

\newcommand{\equilibrium}{E}

\Title{Relativistic Rational Extended Thermodynamics of  Polyatomic Gases with a New  Hierarchy of Moments}

\Author{Takashi Arima$^1$, Maria Cristina Carrisi$^2$, Sebastiano Pennisi$^2$, Tommaso Ruggeri$^3$}

\AuthorNames{Takashi Arima, Maria Cristina Carrisi, Sebastiano Pennisi and Tommaso Ruggeri}

\address{
$^{1}$ \quad  Department of Engineering for Innovation, National Institute of  Technology, Tomakomai College, 
         Tomakomai, Japan; arima@tomakomai-ct.ac.jp\\
  $^{2}$ \quad Department of di Mathematics and Informatics , University of Cagliari,  Cagliari, Italy; mariacri.carrisi@unica.it, spennisi@unica.it \\
$^{3}$ \quad Department of Mathematics and Alma Mater Research Center on Applied Mathematics AM$^2$, University of Bologna, Bologna, Italy; tommaso.ruggeri@unibo.it}

\abstract{A relativistic version of the rational extended thermodynamics of polyatomic gases based on a new hierarchy of moments that  takes into account the total energy composed by the rest energy and the energy of the molecular internal mode is proposed. The moment equations associated with the Boltzmann-Chernikov equation are derived, and the system for the first $15$ equations is closed by the procedure of the maximum entropy principle and by using an appropriate BGK model for the collisional term. The entropy principle with a convex   entropy density is proved in a neighborhood of equilibrium state, and, as a consequence, the system is symmetric hyperbolic and the Cauchy problem is well-posed.
The ultra-relativistic  and classical limits are also studied. The theories with $14$  and $6$ moments are deduced as principal subsystems. Particularly interesting is the subsystem with $6$ fields in which the dissipation is only due to the dynamical pressure. This simplified model can be very useful when bulk viscosity is dominant and might be important in cosmological problems.
Using the Maxwellian iteration, we obtain the parabolic limit, and the heat conductivity, shear viscosity, and bulk viscosity are deduced and plotted.}

\keyword{Relativistic extended thermodynamics; Rarefied polyatomic gas; Causal theory of relativistic fluids}

\begin{document}

\section{Introduction}
Rational extended thermodynamics (RET) is a theory applicable to nonequilibrium phenomena out of local equilibrium. It is expressed by an hyperbolic system of field equations with local constitutive equations and is strictly related to the kinetic theory with the closure method of the hierarchies of moment equations in both classical  and relativistic framework \cite{RET,RS}. 
 
The first relativistic version of the modern RET was given by Liu, M\"uller and Ruggeri (LMR) \cite{LMR} considering the Boltzmann-Chernikov relativistic equation
  \cite{BGK,Synge,KC}:
  \begin{equation}\label{BoltzR}
  p^\alpha \partial_\alpha f = Q,
  \end{equation}
 in which the distribution function  $f$ depends on  $(x^\alpha,  p^\beta)$, where $x^\alpha$ are the space-time coordinates, $p^\alpha$ is the four-momentum, $\partial_{\alpha} = \partial/\partial x^\alpha$, $c$ denotes the light velocity, $m$ is the particle mass in the rest frame, $Q$ is the collisional term and $\alpha, \beta =0,1,2,3$. 
 For monatomic gases, the relativistic moment equations associated with \eqref{BoltzR}, truncated at tensorial index $N+1$, are \footnote{When $n=0$, the tensor reduces to $A^\alpha$. Moreover, the production tensor in the right-side of \eqref{RelmomentMono} is zero for $n=0,1$,  because the first $5$ equations   represent the conservation laws of the particle number and of the energy-momentum, respectively.}:
  \begin{equation}\label{Relmomentseq}
  \partial_\alpha A^{\alpha \alpha_1 \cdots \alpha_n  } =  I^{  \alpha_1 \cdots \alpha_n   }
  \quad \mbox{with} \quad n=0 \, , \,\cdots \, , \,  N 
  \end{equation}
  with  
  \begin{align}\label{RelmomentMono}
  A^{\alpha \alpha_1 \cdots \alpha_n  } = \frac{c}{m^{n-1}} \int_{\R^{3}}
  f  \,  p^\alpha p^{\alpha_1} \cdots p^{\alpha_n}  \, \, d \boldsymbol{P}, \quad I^{\alpha_1 \cdots \alpha_n  } = \frac{c}{m^{n-1}} \int_{\R^{3}}
  Q  \,   p^{\alpha_1} \cdots p^{\alpha_n}  \, \, d \boldsymbol{P} ,
  \end{align}
 and 
  \begin{equation*}
  d \boldsymbol{P} =  \frac{dp^1 \, dp^2 \,
  	dp^3}{p^0} .
  \end{equation*}
  When $N=1$, we have the relativistic Euler system
  \begin{equation}
     \partial_\alpha A^{\alpha } = 0, \quad \partial_\alpha A^{\alpha \beta} =0, \label{Euleros}
     \end{equation}
     where, also in the following,  $A^\alpha \equiv V^\alpha$ and $A^{\alpha\beta} \equiv T^{\alpha\beta}$ have the physical  meaning, respectively, of the particle number vector and the energy-momentum tensor.
   Instead, when $N=2$, we have the LMR theory of a relativistic gas with $14$ fields:
  \begin{equation}
   \partial_\alpha A^{\alpha } = 0, \quad \partial_\alpha A^{\alpha \beta} =0, \quad  \partial_\alpha A^{\alpha \beta \gamma} =  I^{  \beta \gamma  }, \qquad \Big(\gamma=0,1,2,3; \,\, I^\alpha_{\,\,\alpha} =0 \Big). \label{Annals}
   \end{equation}

Recently, Pennisi and Ruggeri first constructed a relativistic ET theory for polyatomic gases with \eqref{Relmomentseq}  in the case of $N=2$ \cite{Annals} (see also \cite{Car1,Car2}) whose moments are given by
\begin{align} 
\begin{split}
& A^\alpha  = m c \inta f p^\alpha \phi(\mathcal{I}) \, d \mathcal{I} \, d \boldsymbol{P}  \, , \\
& A^{\alpha \beta}  = \frac{1}{mc} \inta f p^\alpha p^\beta (mc^2 + \II) \, \phi(\mathcal{I}) \, d \mathcal{I} \, d \boldsymbol{P}  \, , \\
& A^{\alpha \beta \gamma  } = \frac{1}{m^2 c} \int_{\R^{3}}
\int_0^{+\infty} f  \,   p^{\alpha} p^\beta p^{\gamma}  \, \Big( mc^2 + 2\II \Big) \, 
\phi(\mathcal{I}) \, d \mathcal{I} \, d \boldsymbol{P}  \, , 
\end{split}
\label{PS3}
\end{align}
where the distribution function $f(x^\alpha, p^\beta,\mathcal{I})$ depends on the extra energy variable $\mathcal{I}$, similar to the classical one (see \cite{RS} and references therein),  and $\phi(I)$ is the state density of the internal mode.

In \cite{Annals}, by taking the traceless part of the third order tensor, i.e., $A^{\alpha \langle \beta \gamma \rangle}$, as a field instead of $A^{\alpha\beta\gamma}$ in \eqref{Annals}$_3$, the relativistic theory with 14 fields (ET$_{14}$) was proposed. It was also shown that its classical limit coincides with the classical ET$_{14}$ based on the binary hierarchy \cite{Arima-2011,Pavic-2013,RS}. The beauty of the relativistic counterpart is that there exists a single hierarchy of moments, but, as was noticed by the authors, to obtain the classical theory of ET$_{14}$, it was necessary to put the factor 2 in front of $\II$  in the last equation of \eqref{PS3}!
This was also more evident in  the theory with any number of moments where Pennisi and Ruggeri generalized \eqref{PS3} considering the following moments \cite{PRS}:
\begin{align} \label{relRETold}
\begin{split}
& A^{\alpha \alpha_1 \cdots \alpha_n  } = \frac{1}{m^n c} \int_{\R^{3}}
\int_0^{+\infty} f  \,  p^\alpha p^{\alpha_1} \cdots p^{\alpha_n}  \, \Big(mc^2 +  n \II \Big)\, 
\phi(\mathcal{I}) \, d \mathcal{I} \, d \boldsymbol{P}  \, , \\
& I^{\alpha_1 \cdots \alpha_n  } = \frac{1}{m^{n}c} \int_{\R^{3}}
\int_0^{+\infty} Q  \,  p^{\alpha_1} \cdots p^{\alpha_n}  \, \Big( mc^2 +  n\II \Big)\, 
\phi(\mathcal{I}) \, d \mathcal{I} \, d \boldsymbol{P}. \\
\end{split}
\end{align}
In this case, we need a factor $n \II$ in \eqref{relRETold} to obtain, in the classical limit, the binary hierarchy.

To avoid this unphysical situation, Pennisi first noticed that $ (mc^2 +n \II)$ appearing in \eqref{relRETold} are the first  two terms of the Newton binomial formula for $(mc^2 +\mathcal{I})^n/ (mc^2)^{n-1}$
 Therefore he proposed in \cite{Pennisi_2021} to modify, in the relativistic case, the definition of the moments by using the substitution:
 \begin{equation*}
 (mc^2)^{n-1}\Big( mc^2 + n \II \Big) \qquad \text{with \quad } \Big( mc^2 + \II \Big)^n,
 \end{equation*}
 i.e., instead of \eqref{relRETold}, the following moments are proposed:
\begin{align} \label{relRET}
\begin{split}
& A^{\alpha \alpha_1 \cdots \alpha_n  } = \Big(\frac{1}{mc}\Big)^{2n-1} \int_{\R^{3}}
\int_0^{+\infty} f  \,  p^\alpha p^{\alpha_1} \cdots p^{\alpha_n}  \, \Big( mc^2 +  \II \Big)^n\, 
\phi(\mathcal{I}) \, d \mathcal{I} \, d \boldsymbol{P}  \, , \\
& I^{\alpha_1 \cdots \alpha_n  } = \Big(\frac{1}{mc}\Big)^{2n-1} \int_{\R^{3}}
\int_0^{+\infty} Q  \,  p^{\alpha_1} \cdots p^{\alpha_n}  \, \Big( mc^2 +  \II \Big)^n\, 
\phi(\mathcal{I}) \, d \mathcal{I} \, d \boldsymbol{P}. \\
\end{split}
\end{align}
Such definitions are more physical because now the full energy (the sum of the rest frame energy and the energy of internal modes) $mc^2 + \II$ appears in the moments. 

The aim of this paper is to consider the system \eqref{Annals} with moments given by \eqref{relRET}. In this way, for the case with $N=2$ also by taking the trace part of $A^{\alpha\beta\gamma}$ as a field, we have $15$ field equations and to close the system we adopt the molecular procedure of RET  based on the Maximum Entropy Principle. 

The paper is organized as follows. In Section~\ref{sec:momeq}, the values of generic moments in an equilibrium state are estimated in the general case. In Section~\ref{sec:15}, the ET theory for $15$ fields (ET$_{15}$) is proposed, and the constitutive quantities are closed near the equilibrium state. By adopting a variant of BGK  model appropriate for polyatomic gases proposed by Pennisi and Ruggeri \cite{Car1}, the production tensor is derived. In Section~\ref{entro} the four-dimensional entropy flux and the entropy production are deduced within the second order with respect to the nonequilibrium variables. Then, we show the condition of convexity of the entropy density and the positivity of the entropy production which ensure the well-posedness of the Cauchy problem and the entropy principle  as a result. We also discuss  in Section~\ref{diatomico}  the case of  the diatomic gases for which all coefficients are expressed in closed form in terms of the ratio of two Bessel functions similar to the case of monatomic gases. In Section~\ref{sec:ultra}, we study the ultra-relativistic limit. In Section~\ref{sec:sub}, the principal subsystems of ET$_{15}$ are studied. First, we obtain  ET$_{14}$ in which all field variables have physical meaning. Then, at the same level as ET$_{14}$ in the sense of the principal subsystem, there also exists the subsystem with $6$ fields in which the dissipation is only due to the dynamical pressure. This system is important in the case that the bulk viscosity is dominant compared to the shear viscosity and heat conductivity and must be particularly interesting in cosmological problems. The simplest subsystem is the Euler non dissipative case with $5$ fields. In Section~\ref{sec:max}, we use the Maxwellian iteration and, as a result, the phenomenological coefficients of the Eckart theory, that is, the heat conductivity, shear viscosity and bulk viscosity are determined with the present model.  Finally, in Section~\ref{sec:classical}, we show that the classic limit of the present model coincides with the classical ET$_{15}$ studied in \cite{x}.

\section{Distribution function and moments at equilibrium} \label{sec:momeq}
The equilibrium distribution function $f_E$ of polyatomic gas that generalizes the J\"uttner one of monatomic gas
was evaluated in \cite{Annals} with the variational procedure of Maximum Entropy Principle (MEP) \cite{Janes,ET,RET}. Considering the first $5$ balance equations of \eqref{Annals} in equilibrium state:
\begin{equation*} 
A^\alpha_E \equiv  V_E^\alpha =  m \, n U^\alpha,   \quad  \quad
A^{\alpha\beta}_E \equiv T_E^{\alpha \beta} =   p h^{\alpha \beta}   + \frac{e}{c^2} \, U^{\alpha } U^{\beta}.
\end{equation*}
MEP requires that the appropriate distribution function $f\equiv f(x^\alpha,   p^\alpha, 
\mathcal{I})$ is the one which maximizes the entropy density
\begin{equation*}
\rho S = h_E= h^\alpha_E U_\alpha = - k_B \, c \,  U_\alpha \int_{\R^3}
\int_0^{+\infty} f
\ln f p^\alpha \phi(\mathcal{I})  \, d \mathcal{I} \, d \boldsymbol{P} \, ,
\end{equation*}
under the constraints that the temporal part $V^\alpha U_\alpha$ and $T^{\alpha\beta}U_\beta$  are prescribed. 
Here, $k_B, n, \rho(=nm), U^\alpha, h^{\alpha\beta},p,e,S$  are respectively the Boltzmann constant, the  particle number, the mass density, the four-velocity ($U^\alpha U_\alpha= c^2$), the projector tensor $(h^{\alpha\beta}= U^\alpha U^\beta/c^2 - g^{\alpha\beta})$, the pressure, the energy and the entropy density and $g^{\alpha \beta}= \text{diag}(1 \, , \, -1 \, , \, -1 \,, \, -1)$ is the metric tensor.

The equilibrium distribution function  for a rarefied polyatomic gas that maximizes the entropy has the following expression \cite{Annals}:
\begin{equation}\label{5.2n}
{f_E= \frac{n }{\bar{A}(\gamma)} \frac{1}{4 \pi m^3
	c^3} e^{- \frac{1}{k_B T} \left[ \Big( 1 + \frac{\mathcal{I}}{m c^2} \Big) U_\beta
	p^\beta \right]}}, \qquad \bar{A}(\gamma) = \int_0^{+\infty} J_{2,1}^*\, \phi(\mathcal{I})  \, d \, \mathcal{I}
\end{equation}
with $T$ being the absolute temperature,
\begin{equation*}
    \begin{split}
        & J_{m,n}^* = J_{m,n} (\gamma^*), \qquad \gamma^* = \gamma \, \Big( 1
            + \frac{\mathcal{I}}{m \, c^2} \Big), \qquad   \gamma = \frac{m \, c^2}{k_B T},
    \end{split}
\end{equation*}
and
\begin{equation*}
J_{m,n}(\gamma)= \int_0^{+\infty} e^{-\gamma \cosh s}  \sinh^m s \cosh^n s \, d \, s \, ,
\end{equation*}
subjected to the following recurrence relations \cite{LMR,Annals}:
\begin{equation}\label{R}
    \begin{split}
      J_{m+2,n}(\gamma)= J_{m,n+2}(\gamma) - J_{m,n} (\gamma) \, ,
    \end{split}
\end{equation}
\begin{equation}\label{Rbis}
    \begin{split}
        -\gamma J_{m+2,n}(\gamma)= n J_{m,n-1}(\gamma) - (n+m+1) J_{m,n+1} (\gamma) \, .
    \end{split}
\end{equation}
 The pressure and the energy 
 compatible with the equilibrium distribution function \eqref{5.2n} are \cite{Annals}:
\begin{align}\label{10}
    \begin{split}
    & p =\frac{ k_B}{m} \, \rho  T \, , \qquad \qquad e=
    \rho  c^2   \omega(\gamma), \\
    & \text{with } \quad \omega(\gamma)= \frac{\int_0^{+\infty} J_{2,2}^* \, \Big( 1 + \frac{\mathcal{I}}{m c^2} \Big) \, \phi(\mathcal{I})  \, d \, \mathcal{I}}{\int_0^{+\infty} J_{2,1}^* \,  \phi(\mathcal{I})  \, d \, \mathcal{I}} .
    \end{split}
\end{align}
Taking into account that 
$
e = \rho c^2 + \rho \varepsilon, 
$
where $\varepsilon$ is the internal energy, we deduce from \eqref{10}:
\begin{equation}\label{interenergy2}
\varepsilon = c^2(\omega -1).
\end{equation}
Therefore the internal energy  is a function only of $\gamma$ or, it is the same, of $T$ as in the classical case for rarefied gases.

The moments in equilibrium state $A^{\alpha \alpha_1 \cdots \alpha_j  }_E$  for $j \geq 2$ were deduced in \cite{Pennisi_2021}:
\begin{align}\label{11}
    A^{\alpha_1 \cdots \alpha_{j+1 } }_E= \sum_{k=0}^{\left[ \frac{j+1}{2} \right]} \rho c^{2k} \theta_{k,j} \, h^{( \alpha_1 \alpha_2 } \cdots h^{ \alpha_{2k-1} \alpha_{2k} } U^{\alpha_{2k+1} } \cdots U^{\alpha_{j+1} ) } \, ,
\end{align}
where
\begin{align}\label{11b}
    \theta_{k,j} = \frac{1}{2k+1}  \begin{pmatrix}
        j+1 \\ 2k
    \end{pmatrix} \frac{\int_0^{+\infty} J_{2k+2,j+1-2k}^* \, \Big( 1 + \frac{\mathcal{I}}{m c^2} \Big)^j \, \phi(\mathcal{I})  \, d \, \mathcal{I}}{\int_0^{+\infty} J_{2,1}^* \,  \phi(\mathcal{I})  \, d \, \mathcal{I}} \,
\end{align}
are dimensionless functions depending only on $\gamma$.  
Taking into account \eqref{10} and \eqref{11b}, we obtain $\theta_{0,0} =1, \, \theta_{0,1} =\omega(\gamma)$ and 
using the recurrence formula \eqref{R} and \eqref{Rbis}, in \cite{Pennisi_2021}, the following recurrence relations hold:
\begin{align}\label{12b}
    \begin{split}
        & \theta_{0,0}=1 \, ,\\
        & \theta_{0,j+1} = \omega(\gamma) \, \theta_{0,j}  \, - \, \theta^{\, \prime}_{0,j} \hspace{4.6 cm} \mbox{with } \quad ^\prime = \frac{d}{d \gamma},  \\
        & \\
        &  \theta_{h,j+1} = \frac{j+2}{\gamma} \Big( \theta_{h,j} + \frac{j+3-2h}{2h} \theta_{h-1,j} \Big) \hspace{1.9 cm} \mbox{for } h=1, \, \cdots , \,\left[ \frac{j+1}{2} \right] \, , \\
        & \\
        &  \theta_{\frac{j+2}{2},j+1} = \frac{1}{\gamma} \theta_{\frac{j}{2}, \, j} \hspace{6.3 cm} \mbox{for $j$ even}  \, .
    \end{split}
\end{align}
It is interesting to see that all the scalar coefficients can be expressed in terms of the function $\omega(\gamma)$   and of its derivatives with respect to $\gamma$ (or with respect to the temperature $T$) and $\omega$ is strictly related to the internal energy $\varepsilon$ by \eqref{interenergy2}. A similar situation is studied in the article \cite{x} for the non relativistic case.\\
The  values of $\theta_{h,j}$  can be determined, by using the recurrence formula \eqref{12b}, according to the following diagram:
\begin{align*}
\begin{matrix} \theta_{0,0}  & \Rightarrow & \theta_{0,1}  &  \Rightarrow & \theta_{0,2} &  \Rightarrow &  \theta_{0,3} & \cdots \\
        &\searrow &&& \\
        ~  &  & \theta_{1,1}  & \rightarrow & \theta_{1,2}  & \rightarrow & \theta_{1,3} & \cdots &      \\
        & & &  & & \searrow \\
        ~  &  & ~  &   & ~ & &  \theta_{2,3} &  \rightarrow & \theta_{2,4} &\cdots \\
        &
    \end{matrix}
\end{align*}
We see that all the $\theta_{0,j}$ can be obtained from $\theta_{0,0}$ by using eq. $\eqref{12b}_2$ and the other $\theta_{h,j}$ with $j\geq h$ can be obtained from eqs. $\eqref{12b}_{3,4}$. In particular, we can evaluate the following ones that are needed to know for the model with $15$ fields in the subsequent sections:
\begin{align}\label{thetas}
\begin{split}
& \theta_{0,0}=1, \qquad \theta_{0,1}=\omega , \qquad \theta_{0,2}=\omega
   ^2-\omega ',
   \\
   &\theta_{0,3}=\omega ^3+\omega
   ''-3 \omega  \omega ', \qquad \theta_{0,4}=\omega
   ^4-\omega {'''}+4 \omega 
   \omega ''+3 \omega
   '^2-6 \omega ^2
   \omega ' , \\
   & \theta_{1,1}=\frac{1}{\gamma
      }, \qquad \theta_{1,2}=\frac{3}{\gamma ^2} (\gamma  \omega
      +1), \qquad  \theta_{1,3}=\frac{6}{\gamma
            ^3}
      \left[\gamma ^2 (\omega
      ^2-
       \omega ')+2(
      \gamma  \omega
      +1)\right],
      \\
  &     \theta_{1,4}=\frac{10} {\gamma^4 }
  \left\{3
      \gamma  \left[\omega 
      (\gamma  \omega +2)-\gamma 
      \omega '\right]+6+ {\gamma
      ^3}( \omega^3 +\omega ''-3
      \omega  \omega
      ')\right\},\\
      & \theta_{2,3}=\frac{3} {\gamma ^3}
      (\gamma  \omega
         +1), \qquad  \theta_{2,4}=\frac{15}{\gamma
                  ^4}
         \left[\gamma ^2 (\omega
         ^2- \omega ')+3(
         \gamma  \omega
         +1)\right].
\end{split}
\end{align}

\section{The closure for the 15 moments model}\label{sec:15}
In this section, we consider the simplest and physical case, that is, the system \eqref{Relmomentseq} for $n=0,1,2$ with the moments given by \eqref{relRET}:
 \begin{equation}\label{Annalis}
   \partial_\alpha V^{\alpha } = 0, \quad \partial_\alpha T^{\alpha \beta} =0, \quad  \partial_\alpha A^{\alpha \beta \gamma} =  I^{  \beta \gamma  }, \qquad \left(\beta,\gamma=0,1,2,3\right). 
   \end{equation}
with
\begin{align} \label{relRETpol}
\begin{split}
& V^{\alpha } = {mc}  \int_{\R^{3}}
\int_0^{+\infty} f  \,  p^\alpha  \, 
\phi(\mathcal{I}) \, d \mathcal{I} \, d \boldsymbol{P}  \, , \qquad 
 T^{\alpha \beta} = c\int_{\R^{3}}
\int_0^{+\infty} f  \,  p^\alpha p^{\beta}  \, \Big( 1 +  \frac{\II }{m c^2}\Big)\, 
\phi(\mathcal{I}) \, d \mathcal{I} \, d \boldsymbol{P}  \, , \\
& A^{\alpha \beta\gamma } = \frac{c}{m}  \int_{\R^{3}}
\int_0^{+\infty} f  \,  p^\alpha p^{\beta}  p^{\gamma}  \, \Big( 1 +  \frac{\II}{mc^2} \Big)^2\, 
\phi(\mathcal{I}) \, d \mathcal{I} \, d \boldsymbol{P}  \, , \\
& I^{ \beta \gamma } = \frac{c}{m}\int_{\R^{3}}
\int_0^{+\infty} Q  \,  p^{\beta}  p^{\gamma}\, \left(1 +  \frac{\II}{mc^2} \right)^2\, 
\phi(\mathcal{I}) \, d \mathcal{I} \, d \boldsymbol{P}. \\
\end{split}
\end{align}
To close the system \eqref{relRETpol}, we adopt the MEP which requires to find the distribution function that maximizes the non-equilibrium entropy density:
 \begin{equation}\label{entropy}
h= h^\alpha U_\alpha =  - k_B \, c \,  U_\alpha \int_{\R^3}
\int_0^{+\infty} f
\ln f p^\alpha \phi(\mathcal{I})  \,  d\mathcal{I}  \,  d\boldsymbol{P} , \quad \rightarrow \quad \max
 \end{equation}
under the constraints  that the temporal part $V^\alpha U_\alpha, T^{\alpha\beta}U_\alpha$ and  $A^{\alpha\beta\gamma}U_\alpha$  are prescribed. Proceeding in the usual way as indicates in previous papers of RET (see \cite{RS,Annals}), we obtain:
%
%
\begin{equation}\label{f15}
    f_{15}= e^{   -1 -  \frac{\chi}{k_B}}   \, , \quad \mbox{with} \quad
    \chi = m \, \lambda \, + \,  \lambda_{\mu} \, p^{\mu} \, \left( 1 + \, \frac{\mathcal{I}}{m \, c^2} \right) \, + \,  \frac{1}{m} \, \lambda_{\mu \nu} \, p^{\mu} p^{\nu}  \, \left( 1 + \, \frac{\mathcal{I}}{m \, c^2} \right)^2 \, ,
\end{equation}
where $\lambda, \lambda_{\mu}, \lambda_{\mu \nu}$ are the Lagrange multipliers.

Hereafter, recalling the following decomposition of the particle number vector and the energy-momentum tensor
\begin{align}\label{19}
    V^\alpha =\rho  U^\alpha \, , \quad  T^{\alpha \beta} = \frac{e}{c^2} \,  U^{\alpha } U^\beta + \, \left(p \, + \, \Pi\right)
    h^{\alpha \beta} + \frac{1}{c^2} ( U^\alpha  q^\beta +U^\beta  q^\alpha)+   t^{<\alpha \beta>_3} \, ,
\end{align}
we can choose as fields, as usual, $14$ physical variables;  $\rho$, $T$, $U^\alpha$, $\Pi$, $q^\alpha$, $t^{<\alpha \beta>_3}$, where $\Pi$ is the dynamic pressure, $q^\alpha= -h^\alpha_\mu
U_\nu T^{\mu \nu}$ is the heat flux and $t^{<\alpha \beta>_3} = T^{\mu\nu} \left(h^\alpha_\mu h^\beta_\nu - \frac{1}{3}h^{\alpha\beta}h_{\mu\nu}\right)$ is the deviatoric shear viscous stress tensor.  We also recall the constraints:
\begin{equation*}
U^\alpha U_\alpha = c^2, \quad q^\alpha U_\alpha = 0, \quad t^{<\alpha \beta>_3} U_\alpha = 0, \quad t^{<\alpha}_{\,\,\,\,\,\ \alpha >_3} =0,
\end{equation*}
and we choose as the $15$th variable:
    \begin{align}\label{deltina}
\Delta = \frac{4}{c^2} \,  U_\alpha U_\beta U_\gamma \,  \left( A^{\alpha \beta \gamma} \, - \,  A^{\alpha \beta \gamma}_E\right).
    \end{align}
    The pressure $p$ and the energy $e$  as function of $(\rho,T)$ are given in \eqref{10}.
    
\smallskip

{\bf Remark 1.} : 
	For any symmetric tensor $M^{\alpha \beta}$, we can define its traceless part  $M^{< \alpha \beta >}$ and its 3-dimensional traceless part  $M^{< \alpha \beta >_3}$ that is the traceless part of its projection in the 3-dimensional space orthogonal to $U^\alpha$	as follows
	\begin{align*}
	\begin{split}
	&  M^{< \alpha \beta >} = \left( g_\mu^{\alpha} \,  g_\nu^{ \beta} - \, \frac{1}{4} \, g^{\alpha \beta} g_{\mu \nu} \right) \, M^{\mu \nu} = M^{\alpha \beta}  - \, \frac{1}{4} \,  g_{\mu \nu}  \, M^{\mu \nu} g^{\alpha \beta} \, , \\
	& M^{< \alpha \beta >_3} = \left( h_\mu^{\alpha} \,  h_\nu^{ \beta} - \, \frac{1}{3} \, h^{\alpha \beta} h_{\mu \nu} \right) \, M^{\mu \nu}  \, ,
	\end{split}
	\end{align*}
	which are different except for the case in which 
	$M^{\mu \nu} U_\mu=0$, and $M^{\mu \nu} g_{\mu \nu}=0 $. In fact, these conditions indicate that    
	\begin{align*}
	M^{< \alpha \beta >}= M^{< \alpha \beta >_3} \, . 
	\end{align*}
	Moreover in the following  a parenthesis between two indexes indicate the symmetric part.
	
\subsection{The linear deviation from equilibrium}
The thermodynamical definition of the equilibrium according with M\"uller and Ruggeri \cite{RET} is  
the state in  which the entropy production
 vanishes and hence attains its minimum
value. Using this definition, the theorem was proved \cite{BoillatRuggeri-1998,BoillatRuggeriARMA} that the components of the Lagrange multipliers of the balance laws of nonequilibrium variables vanish, and only 
the five Lagrange multipliers corresponding to the equilibrium  conservation laws (Euler System) remain.  In the present case, we have:
\begin{align}\label{mainE}
\begin{split}
& {\lambda}_E = - \frac{1}{T}\left(g+ c^2\right), \quad {\lambda}_{\mu_E}= \frac{U_\mu}{T},  \quad \lambda_{\mu \nu_E} =0,
\end{split} 
\end{align}
where $g =\varepsilon +p/\rho -T S $ is the equilibrium chemical potential. We remark that ${\lambda}_E , {\lambda}_{\mu_E}$   are the components of the \emph{main field} that symmetrize the relativistic Euler system as was first proved by Ruggeri and Strumia  (see \cite{RugStr}).

In the molecular ET approach,  we consider, as usual, the processes near equilibrium.  For this reason, we expand \eqref{f15} around an equilibrium state as follows:
\begin{align}\label{fgenE}
 \begin{split}
 &f_{15} \simeq f_E\Big(1-\frac{1}{k_B}\tilde{\chi}\Big), \\
 &\tilde{\chi}   = m \, (\lambda - \lambda_E) \, + \,  (\lambda_{\mu}-\lambda_{\mu_E}) \, p^{\mu} \, \Big( 1 + \, \frac{\mathcal{I}}{m \, c^2} \Big) \, + \,  \frac{1}{m} \, \lambda_{\mu \nu} \, p^{\mu} p^{\nu}  \, \Big( 1 + \, \frac{\mathcal{I}}{m \, c^2} \Big)^2 \, .
 \end{split}
\end{align}

Inserting the distribution function \eqref{fgenE} into the moments \eqref{relRETpol}, we obtain the following system:
\begin{align}\label{18}
 \begin{split}
    & 0= V^\alpha - V^\alpha_E = - \frac{m}{k_B}\left[ V^\alpha_E (\lambda - \lambda_E) + T^{\alpha \mu}_E \Big(
    \lambda_\mu - \lambda_{\mu_E} \Big) + A^{\alpha \mu \nu}_E \lambda_{\mu \nu} \right] \, , \\
    &  t^{<\alpha \beta>_3}  +\Pi
    h^{\alpha \beta} + \frac{2}{c^2} \, U^{(\alpha } q^{\beta)}  = - \frac{m}{k_B} \left[ T^{\alpha \beta}_E (\lambda - \lambda_E) +  A^{\alpha \beta \mu}_{E} \Big(
    \lambda_\mu - \lambda_{\mu_E} \Big) +   A^{\alpha \beta \mu \nu}_{E} \lambda_{\mu \nu} \right] \, ,  \\
    & A^{\alpha \beta \gamma} - A^{\alpha \beta \gamma}_E = - \frac{m}{k_B} \left[ A^{\alpha \beta \gamma}_E (\lambda - \lambda_E) +  A^{\alpha \beta \gamma \mu}_{E}
    \Big( \lambda_\mu - \lambda_{\mu_E} \Big) +  A^{\alpha \beta \gamma \mu \nu}_{E}
    \lambda_{\mu \nu} \right] \, ,
 \end{split}
\end{align}
where the equilibrium values of the tensors $A^{\alpha \beta \mu }_E, A^{\alpha \beta \mu \nu}_E$ and $A^{\alpha \beta \mu \nu \gamma}_E$ can be obtained by \eqref{11} taking $j=2,3,4$:
\begin{align}\label{A1w}
\begin{split}
& A^{\alpha \beta \gamma}_E= \rho \, \theta_{0,2} \, U^{\alpha} U^{\beta} U^{\gamma} \, + \,
  \rho c^2 \, \theta_{1,2}  \, h^{( \alpha \beta } U^{\gamma ) } , \\
 &   A^{\alpha \beta \mu \nu}_E=  \rho \, \theta_{0,3} \, U^\alpha U^\beta U^\mu U^\nu \, +
    \, \rho \, c^2 \, \theta_{1,3}\, h^{( \alpha \beta } U^\mu U^{\nu ) } \, + \, \rho \,
    c^4 \, \theta_{2,3} \, h^{( \alpha \beta } h^{\mu \nu ) } , \\
&A^{\alpha \beta \gamma \mu \nu}_E=   \rho \, \theta_{0,4} \, U^\alpha U^\beta U^\gamma
U^\mu U^\nu \, + \, \rho \, c^2 \theta_{1,4} \, h^{( \alpha \beta } U^\gamma   U^\mu
U^{\nu ) } \, + \, \rho \, c^4 \theta_{2,4} \, h^{( \alpha \beta } h^{\gamma \mu}
U^{\nu )}\, ,
\end{split}
\end{align}
with the $\theta$'s given in \eqref{thetas}.

The system \eqref{18} permits to deduce the $15$ Lagrange multipliers in terms of the $15$ field variables including $\Delta$ given in \eqref{deltina},  and then we can obtain the remaining  part of the tensor $A^{\alpha \beta \gamma}$.

To solve this system, we consider first eq. \eqref{18}$_{1}$ contracted with $U_\alpha$, eq. \eqref{18}$_{2}$ contracted  with $U_\alpha \, U_\beta$,  eq. \eqref{18}$_{3}$ contracted with $U_\alpha U_\beta U_\gamma / c^3$, eq. \eqref{18}$_{2}$ contracted  with
$h_{\alpha \beta}/3$ and \eqref{18}$_{3}$ contracted with  $U_\alpha h_{\beta \gamma} /(3\, c^ 2)$, obtaining the system
\begin{align}\label{19b}
    \begin{split}
        & \theta_{0,0} \, (\lambda - \lambda_E) + \theta_{0,1} \,  U^\mu \Big(
        \lambda_\mu - \frac{U_\mu}{T} \Big) + \theta_{0,2} \,  U^\mu U^\nu \lambda_{\mu \nu} \,   + \frac{c^2}{3} \theta_{1,2} \,  h^{\mu \nu} \lambda_{\mu \nu} = 0 \, ,  \\
        & \theta_{0,1} \,  (\lambda - \lambda_E) + \theta_{0,2} \,  U^\mu \Big(
        \lambda_\mu - \frac{U_\mu}{T} \Big) + \, \theta_{0,3} \,  U^\mu U^\nu \lambda_{\mu \nu} \, + \frac{c^2}{6} \, \theta_{1,3} \, h^{\mu \nu} \lambda_{\mu \nu}=   0 \, , \\
        & \theta_{0,2} \, (\lambda - \lambda_E) + \, \theta_{0,3} \, U^\mu \Big( \lambda_\mu - \frac{U_\mu}{T}
        \Big) + \, \theta_{0,4} \,  U^\mu U^\nu \lambda_{\mu \nu} \, + \frac{c^2}{10} \, \theta_{1,4} \, h^{\mu \nu} \lambda_{\mu \nu}=  - \, \frac{k_B}{4 \, m^2 \, n \, c^4} \, \Delta \, , \\
        & \theta_{1,1}\,  (\lambda - \lambda_E) + \frac{1}{3} \theta_{1,2} \,  U^\mu \Big( \lambda_\mu - \frac{U_\mu}{T}
\Big) + \, \frac{1}{6} \theta_{1,3} \,  U^\mu U^\nu \lambda_{\mu \nu} + \,   \frac{5}{9} c^2 \theta_{2,3} \,  h^{\mu \nu} \lambda_{\mu \nu}=  - \, \frac{k_B}{m^2  n c^2} \, \Pi \, , \\
        &  \frac{1}{3} \theta_{1,2} \, (\lambda - \lambda_E) + \,  \frac{1}{6} \,  \theta_{1,3} \, U^\mu \Big( \lambda_\mu - \frac{U_\mu}{T}
        \Big) + \,  \frac{1}{10} \, \theta_{1,4} \, U^\mu U^\nu \lambda_{\mu \nu} \, + \frac{c^2}{9} \,  \theta_{2,4} \, h^{\mu \nu} \lambda_{\mu \nu}= \\
&   \quad \quad \quad \quad \quad \quad \quad \quad \quad \quad \quad \quad \quad \quad \quad \quad \quad \quad \quad \quad     = - \, \frac{k_B}{3m^2  c^4 n} \, \Big( A^{\alpha \beta \gamma} - A^{\alpha \beta \gamma}_E \Big) U_\alpha h_{\beta \gamma} \, .
    \end{split}
\end{align}
This is a system of 5 equations  in the 4 unknowns $\lambda - \lambda_E$, $ U^\mu \Big(
\lambda_\mu - \frac{U_\mu}{T} \Big) $, $U^\mu U^\nu \lambda_{\mu \nu}$, $h^{\mu \nu} \lambda_{\mu \nu}$; in order to have solutions, the determinant of the complete matrix must be zero, i.e.,
\begin{align}\label{determ}
    0 = \left|
    \begin{matrix} \theta_{0,0}  & \theta_{0,1}  & \theta_{0,2}  & \frac{1}{3}\theta_{1,2}  & 0  \\
        &&&& \\
        \theta_{0,1}  & \theta_{0,2}  & \theta_{0,3}  & \frac{1}{6} \, \theta_{1,3} & 0  \\
        &&&& \\
        \theta_{0,2}  & \theta_{0,3}  & \theta_{0,4}  & \frac{1}{10} \, \theta_{1,4}  &  - \, \frac{k_B}{4m c^4} \, \Delta  \\
        &&&& \\
        \theta_{1,1}   & \frac{1}{3} \, \theta_{1,2}  & \frac{1}{6} \, \theta_{1,3} & \frac{5}{9} \,   \theta_{2,3}  & - \, \frac{k_B}{m c^2} \, \Pi \\
        &&&& \\
        \frac{1}{3} \, \theta_{1,2}  & \frac{1}{6} \, \theta_{1,3}  & \frac{1}{10} \, \theta_{1,4}  & \frac{1}{9} \, \theta_{2,4}  & - \, \frac{k_B}{3m \, c^4} \, \Big( A^{\alpha \beta \gamma} - A^{\alpha \beta \gamma}_E \Big) U_\alpha h_{\beta \gamma}
    \end{matrix}
    \right| \, .
\end{align}
By defining
\begin{align*}
& D_4 =  \left|
\begin{matrix} \theta_{0,0}  & \theta_{0,1}  & \theta_{0,2}  & \frac{1}{3}\theta_{1,2}   \\
    &&&& \\
    \theta_{0,1}  & \theta_{0,2}  & \theta_{0,3}  & \frac{1}{6} \, \theta_{1,3}   \\
    &&&& \\
    \theta_{0,2}  & \theta_{0,3}  & \theta_{0,4}  & \frac{1}{10} \, \theta_{1,4}   \\
    &&&& \\
    \theta_{1,1}   & \frac{1}{3} \, \theta_{1,2}  & \frac{1}{6} \, \theta_{1,3} & \frac{5}{9} \,   \theta_{2,3}
\end{matrix}
\right| \, ,  \nonumber \\
\nonumber \\
& N^\Pi    = - \, \left|
\begin{matrix} \theta_{0,0}  & \theta_{0,1}  & \theta_{0,2}  & \frac{1}{3}\theta_{1,2}   \\
    &&&& \\
    \theta_{0,1}  & \theta_{0,2}  & \theta_{0,3}  & \frac{1}{6} \, \theta_{1,3}   \\
    &&&& \\
    \theta_{0,2}  & \theta_{0,3}  & \theta_{0,4}  & \frac{1}{10} \, \theta_{1,4}   \\
    &&&& \\
    \frac{1}{3} \, \theta_{1,2}  & \frac{1}{6} \, \theta_{1,3}  & \frac{1}{10} \, \theta_{1,4}  & \frac{1}{9} \, \theta_{2,4}
\end{matrix}
\right| \, ,  
    \quad 
    N^\Delta =  \left|
    \begin{matrix} \theta_{0,0}  & \theta_{0,1}  & \theta_{0,2}  & \frac{1}{3}\theta_{1,2}   \\
        &&&& \\
        \theta_{0,1}  & \theta_{0,2}  & \theta_{0,3}  & \frac{1}{6} \, \theta_{1,3}   \\
        &&&& \\
        \theta_{1,1}   & \frac{1}{3} \, \theta_{1,2}  & \frac{1}{6} \, \theta_{1,3} & \frac{5}{9} \,   \theta_{2,3}   \\
        &&&& \\
        \frac{1}{3} \, \theta_{1,2}  & \frac{1}{6} \, \theta_{1,3}  & \frac{1}{10} \, \theta_{1,4}  & \frac{1}{9} \, \theta_{2,4}
    \end{matrix}
    \right| \, ,
\end{align*}
the \eqref{determ} gives:
\begin{align}\label{21}
\frac{1}{3 \, c^2} \, \left( A^{\alpha \beta \gamma} - A^{\alpha \beta \gamma}_E \right) U_\alpha h_{\beta \gamma} = - \, \frac{N^\Pi}{D_4} \, \Pi \, - \, \frac{N^\Delta}{D_4} \frac{1}{4c^2} \, \Delta \, .
\end{align}

We contract now eq. \eqref{18}$_{1}$  with $h_\alpha^\delta$, eq. \eqref{18}$_{2}$ with $U_\alpha \, h_\beta^\delta$,  eq. \eqref{18}$_{3}$  with  $U_\alpha U_\beta h_\gamma^\delta/c^3$   and \eqref{18}$_{3}$  with  $h_\alpha^\delta h_{\beta \gamma}/(3\, c^ 2)$,  obtaining the system
\begin{align}\label{21bis}
\begin{split}
& c^2 \theta_{1,1} \, h^{\delta \mu}(\lambda_\mu - \lambda_{\mu_E}) +\frac{2}{3}  c^2 \theta_{1,2} \,
 \, U^\mu h^{\delta \nu} \lambda_{\mu \nu} = 0 \, ,  \\
& c^2 \theta_{1,2} \, h^{\delta \mu}(\lambda_\mu - \lambda_{\mu_E}) +\, c^2 \theta_{1,3} \,
 \, U^\mu h^{\delta \nu} \lambda_{\mu \nu}  = - \frac{3 \, k_B}{m^2 c^2 n} \, q^\delta \, ,  \\
& c^2 \theta_{1,3} \, h^{\delta \mu}(\lambda_\mu - \lambda_{\mu_E}) + \frac{18}{15} \, c^2 \theta_{1,4} \,
  \, U^\mu h^{\delta \nu} \lambda_{\mu \nu}  = \frac{6 \, k_B}{m^2 c^4 n} \, \left( A^{\alpha \beta \gamma} - A^{\alpha \beta \gamma}_E \right) U_\alpha U_\beta h_\gamma^\delta \, ,  \\
& \frac{5}{3} \, c^4 \theta_{2,3} \, h^{\delta \mu}(\lambda_\mu - \lambda_{\mu_E}) + \frac{2}{3} c^4 \theta_{2,4}\,
   \, U^\mu h^{\delta \nu} \lambda_{\mu \nu} = \frac{k_B}{m^2 n} \, \left( A^{\alpha \beta \gamma} - A^{\alpha \beta \gamma}_E \right) h_{\alpha \beta} h_\gamma^\delta\, .
\end{split}
\end{align}
By eliminating the parameters $h^{\delta \mu}(\lambda_\mu - \lambda_{\mu_E})$ and $U^\mu h^{\delta \nu} \lambda_{\mu \nu}$ from these equations, we obtain
\begin{align}\label{22}
\begin{split}
& \left( A^{\alpha \beta \gamma} - A^{\alpha \beta \gamma}_E \right) U_\alpha U_\beta h_\gamma^\delta = - \, c^2
 \frac{N_3}{D_3} \, q^\delta \, , \\
 & \left( A^{\alpha \beta \gamma} - A^{\alpha \beta \gamma}_E \right) h_{\alpha \beta} h_\gamma^\delta = - \,
  \frac{N_{31}}{D_3} \, q^\delta \, ,
\end{split}
\end{align}
with
\begin{align*}
D_3 = \left|
\begin{matrix}  \theta_{1,1}  &  \theta_{1,2}   \\
& \\
\theta_{1,2}  & \frac{3}{2} \, \theta_{1,3}  \,
\end{matrix}
\right| \, , \quad N_3 =  \frac{1}{2} \, \left|
\begin{matrix} \theta_{1,1}  & \theta_{1,2}   \\
    & \\
    \theta_{1,3}  & \frac{9}{5} \,  \theta_{1,4}   \,
\end{matrix}
\right| \, ,  \quad   N_{31} =  \left|
\begin{matrix}  \theta_{1,1}  &  \theta_{1,2}  \\
    & \\
    5 \, \theta_{2,3}  & 3 \, \theta_{2,4}   \,
\end{matrix}
\right| \, .
\end{align*}
We contract now  eq. \eqref{18}$_{2}$  with $h_\alpha^{< \delta } \, h_\beta^{\theta >_3 }$   and \eqref{18}$_{3}$  with  $h_\alpha^{< \delta } \, h_\beta^{\theta >_3 } U_\gamma$, obtaining
\begin{align}\label{22b}
\begin{split}
& - \, \frac{k_B}{m} \, t^{<\delta \theta >_3} = \frac{2}{3} \, mn c^4 \theta_{2,3} \,  h^{\mu < \delta }  h^{\theta >_3  \nu} \lambda_{\mu \nu} \, ,  \\
& \left( A^{\alpha \beta \gamma} - A^{\alpha \beta \gamma}_E \right) h_\alpha^{< \delta } \, h_\beta^{\theta >_3} U_\gamma = - \, \frac{2}{15} \, \frac{m}{k_B} \, mn \, c^6 \, \theta_{2,4} \,  h^{\mu < \delta }  h^{\theta >_3  \nu} \lambda_{\mu \nu} \, ,
\end{split}
\end{align}
from which it follows
\begin{align}\label{23}
\left( A^{\alpha \beta \gamma} - A^{\alpha \beta \gamma}_E \right) h_\alpha^{< \delta } \, h_\beta^{\theta >_3 } U_\gamma = C_5  \, c^2 \, t^{<\delta \theta >_3} \,  \qquad \mbox{with} \qquad C_5 = \frac{1}{5} \, \frac{\theta_{2,4}}{\theta_{2,3}} \, .
\end{align}
Finally, \eqref{18}$_{3}$ contracted with $h_\alpha^{< \delta} \,  h_\beta^\theta \, h_\gamma^{\psi>_3 }$ gives
\begin{align*}
    \Big( A^{\alpha \beta \gamma} - A^{\alpha \beta \gamma}_E \Big)h_\alpha^{< \delta} \,  h_\beta^\theta \, h_\gamma^{\psi>_3 } = 0 \, .
\end{align*}
This result, jointly with \eqref{21}, \eqref{22}, \eqref{23}, gives the decomposition of the triple tensor $A^{\alpha \beta \gamma}$:
\begin{align*}
    \begin{split}
 A^{\alpha \beta \gamma} - A_E^{\alpha \beta \gamma} & = \frac{1}{4c^4} \, \Delta \, U^\alpha U^\beta U^\gamma - \, \frac{3}{4c^2}\, \frac{N^\Delta}{D_4} \, \Delta \, h^{(\alpha \beta} U^{\gamma)} - 3 \, \frac{N^\Pi}{D_4} \, \Pi h^{(\alpha \beta} U^{\gamma)} \\
        &   + \frac{3}{c^2}
        \frac{N_3}{D_3}  \, q^{(\alpha} U^\beta U^{\gamma)} + \frac{3}{5} \frac{N_{31}}{D_3} h^{(\alpha
            \beta} q^{\gamma)} + 3  C_5 t^{(<\alpha \beta >_3} U^{\gamma)}
        \, .
    \end{split}
\end{align*}
Thanks to eq. \eqref{A1w}$_1$, we have the closure of the triple tensor in terms of the physical variables:
\begin{align}\label{24}
    \begin{split}
        A^{\alpha \beta \gamma} &= \left( \rho\, \theta_{0,2}  + \, \frac{1}{4c^4} \, \Delta \right)  U^{\alpha } U^{\beta} U^\gamma + \left( \rho \, c^2\,\theta_{1,2}  - \, \frac{3}{4c^2}\, \frac{N^\Delta}{D_4} \, \Delta \, - 3 \, \frac{N^\Pi}{D_4} \, \Pi \right)
        \,  h^{(\alpha\beta} U^{\gamma)} \\
        & + \frac{3}{c^2} \frac{N_3}{D_3}  \, q^{(\alpha} U^\beta U^{\gamma)} + \frac{3}{5} \frac{N_{31}}{D_3} h^{(\alpha
            \beta} q^{\gamma)} + 3  C_5 t^{(<\alpha \beta >_3} U^{\gamma)}
        \, .
    \end{split}
\end{align}

\subsection{Inversion of the Lagrange Multipliers}
In this section, we present the explicit expression of the Lagrange Multipliers in terms of the 15 physical independent variables. 
From the representation theorems, their  are expressed as follows:
\begin{align}\label{RT}
\begin{split}
\lambda - \lambda_E &= a_1 \Pi + a_2 \Delta, \\
\lambda_\mu - \lambda_{\mu_E} &= \left(b_1 \Pi + b_2 \Delta\right)U_\mu + b_3 q_\mu,\\
\lambda_{\mu\nu} &= \left( \alpha_1 \Pi + \beta_1 \Delta\right) U_\mu U_\nu + \left(\alpha_2 \Pi + \beta_2 \Delta\right) h_{\mu\nu} +  \alpha_3 \left(q_{\mu} U_{\nu} +q_{\nu} U_{\mu}\right)  + \alpha_4 t_{<\mu\nu >_3},
\end{split}	
\end{align}
where  $\lambda_E$ and $\lambda_{\mu_E}$ can be found in eq. \eqref{mainE}, and the coefficients $a_{1,2}, b_{1,2,3}$, $\alpha_{1,2,3,4}$ and $\beta_{1,2}$ are  functions of $\rho$ and $\gamma$. By using eqs. \eqref{19b}, \eqref{21bis} and \eqref{22b}, it is possible to obtain the explicit expressions of these coefficients.

For convenience, let us denote by $D_4^{ij}$  the minor determinant obtained from $D_4$ by deleting its $i$th row and $j$th column. From system \eqref{19b}, we  obtain 
\begin{align}\label{EP2}
	\begin{split}
	 \lambda - \lambda_E &= -\frac{k_B}{m c^4 \rho\, D_4} \left(- \Pi \, c^2 D_4^{41}
	+ \frac{\Delta}{4} D_4^{31} \right) , \\
	&\\
	U^\mu(\lambda_\mu-\lambda_{\mu_E}) &= -\frac{k_B}{m c^4 \rho\, D_4} \left( \Pi \, c^2 D_4^{42}
	- \frac{\Delta}{4} D_4^{32} \right), \\
	&\\
	U^\beta U^{\gamma } \lambda_{\beta\gamma} &= -\frac{k_B}{m c^4 \rho\, D_4} \left(- \Pi \, c^2 D_4^{43}
		+ \frac{\Delta}{4} D_4^{33} \right), \\
		&\\
		h^{\beta\gamma}\lambda_{\beta\gamma} &= -\frac{k_B}{m c^4 \rho\, D_4} \left(\Pi D_4^{44} - \frac{\Delta}{4c^2} D_4^{34}
		\right).
	\end{split}
\end{align}
From system \eqref{21bis} we obtain
\begin{align}\label{EP3}
h^{\delta \mu}\left(\lambda_\mu - \lambda_{\mu_E}\right) = \frac{3\, k_B \theta_{1,2}}{m c^4 \rho\, D_3} q^\delta \qquad\qquad  \text{and} \qquad	\qquad U^\beta h^{\gamma\delta}\lambda_{\beta\gamma} = -\frac{9 k_B \theta_{1,1} }{2 m c^4 \rho\, D_3} q^\delta.
\end{align}
Finally, from eq. \eqref{22b} we have
\begin{align*}
	h^{\beta <\delta} h^{\theta >_3  \gamma}\lambda_{\beta\gamma} = -\frac{3 k_B}{2m c^4 \rho\theta_{2,3}} t^{<\delta\theta >_3},
\end{align*}
that, multiplied by $t_{<\delta\theta >_3}$, gives
\begin{align}\label{EP4}
	t^{<\beta \gamma>_3}\lambda_{\beta\gamma} = -\frac{3 k_B}{2m c^4 \rho\theta_{2,3}} t^{<\beta \gamma>_3} t_{<\beta \gamma>_3}.
\end{align}
By comparing eqs. \eqref{RT}$_1$ with \eqref{EP2}$_1$ we have
\begin{align}\label{coef1}
		a_1= \frac{k_B}{m c^2 \rho\, D_4} D_4^{41}, \qquad \qquad
	a_2= -\frac{k_B}{4\,m c^4 \rho\, D_4} D_4^{31} .
\end{align}
By multiplying eq. \eqref{RT}$_2$ respectively times $U^\mu$ and $h^{\mu\delta}$, and using eqs. \eqref{EP2}$_2$ and \eqref{EP3}$_1$ we have
\begin{align}\label{coef2}
	b_1= - \frac{k_B}{m c^4 \rho\, D_4} D_4^{42}, \quad \quad
	b_2= \frac{k_B}{4\,m c^6 \rho\, D_4} D_4^{32},  
 \quad \quad
	b_3 = \frac{3\, k_B \theta_{1,2}}{m c^4 \rho D_3}.
\end{align}
Finally, by multiplying equation \eqref{RT}$_3$ respectively times $U^\mu \, U^\nu$, $h^{\mu\nu}$, $U^\nu h^{\mu\delta}$, $h^{\mu <\delta}h^{\theta> \nu}$ and using eqs. \eqref{EP2}-\eqref{EP4} we obtain that
\begin{align}\label{coef3}
	\begin{split}	
		\alpha_1 &= \frac{k_B}{m c^6 \rho\, D_4} D_4^{43}, \qquad \qquad
		\alpha_2 = -\frac{k_B}{3 m c^4 \rho\, D_4} D_4^{44}, \\
		\alpha_3 &= \frac{9 k_B \theta_{1,1} }{2 m c^6 \rho\, D_3}, \qquad \qquad  \quad
		\alpha_4 =  -\frac{3 k_B}{2m c^4 \rho\theta_{2,3}},\\
		\beta_1 &= -\frac{k_B}{4 m c^8\rho\, D_4} D_4^{33}, \qquad \quad
		\beta_2 = \frac{k_B}{12 m c^6 \rho\, D_4} D_4^{34}.
	\end{split}
\end{align}

\subsection{Production term with a variant BGK model}

To complete the closure  of the system \eqref{Annalis}, we need to have the expression of the production tensor $I^{\beta \gamma}$. It depends on the collisional term $Q$ (see \eqref{relRETpol}$_2$), and obtaining the expression of $Q$ is an hard task in relativity. Usually, for monatomic gas it is adopted the relativistic generalization of the BGK approximation first made by Marle \cite{Mar2,Mar3} and successively by Anderson and Witting \cite{AW}. The  Marle model is an extension of the classical BGK model in the Eckart frame \cite{KC,E}, and the   Anderson-Witting model obtains such extension using the Landau-Lifshitz frame \cite{KC,LL}. There are some weak points for Marle model,  and  the  Anderson-Witting model uses the Landau-Lifshitz four velocity.  
Starting from these considerations, Pennisi and Ruggeri proposed a variant of Anderson-Witting model in the Eckart frame both for monatomic and polyatomic gases, and proved that the conservation laws of particle number and energy-momentum are satisfied and the H-theorem holds \cite{Car1} (see also \cite{RS}). 
In the polyatomic case, the following collision term has been proposed:
\begin{align}\label{P1}
    Q=\frac{U^\alpha p_\alpha}{c^2 \tau}\left(f_E-f-f_Ep^\mu q_\mu \frac{1+\frac{\mathcal{I}}{m c^2}}{bmc^2}\right),
\end{align}
where $3b$ is the coefficient of $h^{(\alpha \beta} U^{\gamma )}$ in eq. \eqref{A1w}$_1$, i.e
$3 b=\rho c^2\theta_{1,2}$.\\
The most general expression of a nonequilibrium double tensor as a linear function of $\Delta$, $\Pi$, $t^{<\mu\nu>_3}$ and $q^{\mu}$ is the following:
\begin{align*}
    I^{\beta \gamma}= (B_1^\Delta \, \Delta  + \, B_1^\Pi \, \Pi ) \, U^\beta U^\gamma \, + ( B_2^\Delta \, \Delta \, + \, B_2^\Pi \, \Pi ) h^{\beta \gamma}  + \, B^q \, U^{(\beta } \, q^{\gamma )} \, + \, B^t \, t^{< \beta \gamma >_3}.
\end{align*}
In order to determine the coefficients in  $I^{\alpha\beta}$, we have to substitute eq. \eqref{P1} into eq. $\eqref{relRETpol}_4$, obtaining
\begin{align*}
    I^{\beta\gamma} &= \frac{c}{m} \int_{\R^{3}}
    \int_0^{+\infty} \frac{U^\alpha p_\alpha}{c^2 \tau}\Big(f_E-f-f_Ep^\mu q_\mu \frac{1+\frac{\mathcal{I}}{m c^2}}{bmc^2}\Big) \,  p^{\beta} p^{\gamma}  \, \Big( 1 + \, \frac{\mathcal{I}}{m \, c^2} \Big)^2 \,
    \phi(\mathcal{I}) \,  \, d \, \mathcal{I}  d\boldsymbol{P}= \nonumber \\
    &= \frac{U_\alpha}{c^2 \tau} (A_E^{\alpha\beta\gamma}- A^{\alpha\beta\gamma}) - 3 \frac{U_\alpha q_\mu}{ \theta_{1,2} m^2 n c^6 \tau} A_E^{\alpha\beta\gamma\mu},
\end{align*}
then we have
\begin{align}\label{P3}
 \begin{split}
    B_1^\Delta &= - \frac{1}{4c^4\tau}, \qquad  B_1^\Pi =0,  \qquad B_2^\Delta =  \frac{1}{4c^2\tau} \, \frac{N^\Delta}{D_4}, \qquad B_2^\Pi = \, \frac{1}{\tau} \, \frac{N^\Pi}{D_4}  \\
    B^q &=  \frac{1}{c^2 \tau} \, \Big( \frac{\theta_{1,3}}{\theta_{1,2}} \, - 2 \frac{N_3}{D_3} \Big) \, , \quad  B^t = - \, \frac{1}{\tau} \, C_5 \, .
 \end{split}
\end{align}
Therefore the final expression of  the production term $I^{\beta\gamma}$ is
\begin{align}\label{P22}
    I^{\beta\gamma} =\frac{1}{\tau}
\left\{    
        - \frac{1}{4c^4  } \Delta  \,  U^\beta U^{\gamma } + \Big( \frac{1}{4c^2} \frac{N^\Delta}{D_4}\Delta +\frac{N^\Pi}{D_4} \Pi \Big) h^{\beta\gamma} + \Big( -\frac{2}{c^2 } \frac{N_3}{D_3} +\frac{\theta_{1,3}}{\theta_{1,2}}\frac{1}{c^2 } \Big)q^{(\beta} U^{ \gamma )} -   C_5 t^{<\beta\gamma>_3}
     \right\}
\end{align}

We summarize the results of  this section as: 
\begin{statement}
The closed system \eqref {Annalis} obtained via MEP is the one for which $V^\alpha, T^{\alpha \beta}, A^{\alpha\beta \gamma}, I^{\beta \gamma} $ are given explicitly in terms of the $ 15 $ fields ($\rho,\gamma,\Pi, \Delta, U^\alpha, q^\alpha, t^{<\alpha \beta>_3}$) using the expressions \eqref {19}, \eqref {24} and \eqref {P22}. All coefficients are completely determined in terms of a single  function  $\omega(\gamma)$  given by eq. \eqref {10}$_3 $ and its derivatives up to the order $ 3 $. Observe, by taking into account \eqref {interenergy2}, that the coefficients $ \theta$'s given in \eqref {thetas} can be formally written in terms of the internal energy $ \varepsilon $ and its derivatives.
\end{statement}
\subsection{Closed system of the field equations and material derivative}
It is possible to write now explicitly the differential system for the field variables using the material derivative.
The relativistic material derivative of a function $f$ is defined as the derivative with respect to the proper time $\bar{\tau}$ along the path of the particle:
\begin{align}\label{matde}
  \dot{f}  =  \frac{d f}{d \bar{\tau}} =  \frac{d f}{dt} \frac{dt}{d\bar{\tau}} = \Gamma (\partial_t f + v^j \partial_j f) = U^\alpha \partial_\alpha f,
  \end{align}
  where $\Gamma$ is the Lorentz factor and we take into account that
  \[
  U^\alpha = \frac{dx^\alpha}{d\bar{\tau}} \equiv (\Gamma c, \Gamma v^j),
  \]
  where $v^j$ is the velocity. 
Now we observe that for any balance law we can have the following identity:
\begin{align*}
    I^{\alpha_1 \cdots \alpha_n} = \partial_\alpha \, A^{\alpha \alpha_1 \cdots \alpha_n} = g^\beta_\alpha \, \partial_\beta \, A^{\alpha \alpha_1 \cdots \alpha_n} =  \Big( -h^\beta_\alpha + \frac{U^\beta U_\alpha}{c^2} \Big) \, \partial_\beta \, A^{\alpha \alpha_1 \cdots \alpha_n} = \\
    = \frac{U_\alpha}{c^2} \, \dot{A}^{\alpha \alpha_1 \cdots \alpha_n} \, - \, h^\beta_\alpha \, \partial_\beta \, A^{\alpha \alpha_1 \cdots \alpha_n} \, .
\end{align*}
In our case with $n=0,1,2$, these equations are written as follows:
\begin{align*}
    \begin{split}
        & \partial_\alpha \left( \rho U^\alpha \right) =0, \quad \, h_{\delta \beta} \, \left( \frac{U_\alpha}{c^2} \, \dot{T}^{\alpha \beta} \, - \, h^\mu_\alpha \, \partial_\mu \, T^{\alpha \beta} \right)=0, \quad \, U_{\beta} \, \left( \frac{U_\alpha}{c^2} \, \dot{T}^{\alpha \beta} \, - \, h^\mu_\alpha \, \partial_\mu \, T^{\alpha \beta} \right)=0, \, , \\
        & h_{\delta \beta} \, h_{\theta \gamma} \, \left( \frac{U_\alpha}{c^2} \, \dot{A}^{\alpha \beta \gamma} \, - \, h^\mu_\alpha \, \partial_\mu \, A^{\alpha \beta \gamma} \, - \, I^{\beta \gamma} \right)=0 \, , \\
        & h_{\delta \beta} \, U_{\gamma} \, \left( \frac{U_\alpha}{c^2} \, \dot{A}^{\alpha \beta \gamma} \, - \, h^\mu_\alpha \, \partial_\mu \, A^{\alpha \beta \gamma} \, - \, I^{\beta \gamma} \right)=0, \quad 
        U_{\beta} \, U_{\gamma} \, \left( \frac{U_\alpha}{c^2} \, \dot{A}^{\alpha \beta \gamma} \, - \, h^\mu_\alpha \, \partial_\mu \, A^{\alpha \beta \gamma} \, - \, I^{\beta \gamma} \right)=0 \, .
    \end{split}
\end{align*}
By using the expressions $\eqref{19}$, \eqref{24} and \eqref{P22}, respectively for $V^\alpha, T^{\alpha \beta}$, $A^{\alpha \beta \gamma}$ and $I^{\beta \gamma}$, we see that these become
{\small
\begin{align}\label{derivmat}
        & \dot{\rho}  + \rho \, \partial_\alpha \, U^\alpha=0 \, ,  \nonumber \\
        & -\frac{e+p+ \Pi}{c^2}\,  \dot{U}^\delta  \, + \, \frac{1}{c^2}\,  h^{\delta}_{\beta} \, \dot{q}^\beta \, + \, \frac{1}{c^2}\,  t^{< \alpha \delta >_3} \, \dot{U}_\alpha \, - \, h^{\delta \mu} \, \partial_\mu (p+ \Pi) \, - \, \frac{1}{c^2}\, q^\mu \, \partial_\mu U^\delta   \, -\frac{1}{c^2}\, q^\delta \, \partial_\alpha U^\alpha \, - \, h^\delta_\beta \, h_\alpha^\mu  \, \partial_\mu \, t^{< \alpha \beta >_3} =0 \, , \nonumber  \\
        & \dot{e} \, + \, 2 \, \frac{U_\alpha}{c^2} \, \dot{q}^\alpha \, + \, (e+p+ \Pi)\,   \partial_\alpha U^\alpha \, - \, h_\alpha^\mu  \, \partial_\mu q^\alpha \, -  t^{< \alpha \beta >_3} \, \partial_\alpha U_\beta =0 \, , \nonumber \\
 & h_{\delta \beta} \,\Big( \frac{1}{3} \rho c^2 \theta_{1,2}  \, - \, \frac{1}{4 \, c^2} \, \frac{N^\Delta}{D_4} \, \Delta \, - \, \frac{N^\Pi}{D_4} \, \Pi \Big)^\bullet
        + \, C_5 \, h_{\delta \gamma} \, h_{\theta \beta} \, \dot{t}~^{< \theta \gamma>_3} \, + \, t_{< \delta \beta >_3} \, \dot{C}_5 - 
        \frac{2}{c^2} \, \Big( \frac{N_3}{D_3} \, + \, \frac{1}{5} \, \frac{N_{31}}{D_3} \Big) q_{( \delta} \, h_{\beta ) \gamma} \, \dot{U}^\gamma \, -  \nonumber \\
      & \hspace{2cm} \frac{1}{5 \, c^2} \, \frac{N_{31}}{D_3} \, h_{\beta \delta} \, q^\alpha \, \dot{U}_\alpha + 
        \, \Big(- \, \frac{1}{3} \rho c^2 \theta_{1,2}  \, + \, \frac{1}{4 \, c^2} \, \frac{N^\Delta}{D_4} \, \Delta \, + \,
        \frac{N^\Pi}{D_4} \, \Pi \Big) \,
        \left[   - h_{\delta \beta} \partial_\alpha \,  U^\alpha \, + \, 2 \,
        h_{\theta ( \delta } \, h_{\beta )}^\mu \, \partial_\mu \, U^\theta \right] \, + \nonumber \\
        & \hspace{2cm}  \frac{1}{5} \, \Big( q^\mu h_{\delta \beta} +  2 \, q_{( \delta} h_{\beta )}^\mu \Big) \, \partial_\mu \, \Big( \frac{N_{31}}{D_3} \Big)
        - \, \frac{1}{5} \, \frac{N_{31}}{D_3} \,
        \left[ h_{\delta \beta} \, h^\mu_\alpha \,  \partial_\mu \, q^\alpha  + 2 \,
        h_{\theta ( \delta} h^\mu_{\beta )} \, \partial_\mu \, q^\theta \right] + \nonumber  \\
        & \hspace{2cm}  C_5 \,\left[ \, t_{< \delta \beta >_3} \,   \partial_\alpha
        \,  U^\alpha  \,+ \, 2 \,  t^{< \mu \gamma >_3} h_{\gamma ( \beta}  \,
         h_{\delta ) \theta} \,  \partial_\mu \, U^\theta  \right] = \nonumber \\
        & \hspace{2cm} \frac{1}{\tau} \, \Big( \frac{1}{4c^2} \, \frac{N^\Delta}{D_4} \,
        \Delta \, + \, \frac{N^\Pi}{D_4} \, \Pi \Big) h_{\delta \beta} \, - \, \frac{1}{\tau}
        \,  C_5 \, t_{< \delta \beta >_3}\, , \\
        &  h_{\beta \delta} \, \dot{U}^\beta \Big( \rho  \theta_{0,2}  c^2 +  \frac{2}{3} \rho c^2 \theta_{1,2}  + \frac{1}{4 \, c^2} \,
        \Delta - \, \frac{1}{2 \, c^2} \, \frac{N^\Delta}{D_4} \, \Delta
        - \, 2 \, \frac{N^\Pi}{D_4} \, \Pi \Big) + 
          h_{\beta \delta} \, \frac{N_3}{D_3} \,  \dot{q}^\beta - \, q_\delta \,  \Big( \frac{N_3}{D_3} \Big)^\bullet \, + \, \Big( 2 \, C_5 \, - 1 \Big) \, t^{< \delta \gamma >_3} \, \dot{U}_\gamma \, - \nonumber \\
        & \hspace{2cm}  h_\delta^\mu \, \partial_\mu \, \Big( \frac{1}{3} \rho c^4 \theta_{1,2}   - \frac{1}{4} \,
        \frac{N^\Delta}{D_4} \, \Delta
        -  \, \frac{N^\Pi}{D_4} \, c^2 \, \Pi \Big) - 
        \Big( \frac{N_3}{D_3} \, + \, \frac{1}{5}  \, \frac{N_{31}}{D_3} \Big) \, \Big( q^\mu \, \partial_\mu \, U_\delta \,+  q_\delta \,  \partial_\alpha \, U^\alpha \Big) + \nonumber \\
        & \hspace{2cm} \frac{1}{5}  \, \frac{N_{31}}{D_3}  \, h^\mu_\delta \, q^\gamma \, \partial_\mu \, U_\gamma \,+ \, h_\alpha^\mu \,  \partial_\mu \, \Big( C_5 \, c^2 \, t^{< \alpha}_{~~~~\delta >_3} \Big) =
        \frac{1}{\tau} \,  \Big( \frac{N_3}{D_3} \, - \, \frac{\theta_{1,3}}{2 \, \theta_{1,2}} \Big) \, q_\delta  \, , \nonumber \\
        &  \Big( \rho  \theta_{0,2}  c^4 +  \frac{1}{4} \, \Delta  \Big)^\bullet - \, 3 \, \frac{N_3}{D_3} \, q^\alpha \, \dot{U}_\alpha \, + 
        \partial_\alpha \, U^\alpha \cdot \Big( \rho  \theta_{0,2}  c^4 +  \frac{2}{3} \rho c^4 \theta_{1,2}   + \frac{1}{4}
        \, \Delta - \, \frac{1}{2} \, \frac{N^\Delta}{D_4} \, \Delta - \, 2 \, \frac{N^\Pi}{D_4} \, \Pi \, c^2 \Big) \, - \nonumber  \\
        & \hspace{2cm}  h_\alpha^\mu \, \partial_\mu \, \Big( \frac{N_3}{D_3} \, c^2 \, q^\alpha \Big) \, - \, 2 \, C_5 c^2 \, t^{< \mu \gamma >_3} \, \partial_\mu U_\gamma =  - \, \frac{1}{4 \, \tau} \, \Delta  \, . \nonumber 
\end{align}
}
\normalsize
It may be useful to decompose \eqref{derivmat}$_{4}$ into the trace and spatial  traceless parts. The trace part is given by
\begin{align}\label{tracciap}
	\begin{split}
		&  \,\Big(  \rho c^2 \theta_{1,2}  \, - \, \frac{3}{4 \, c^2} \, \frac{N^\Delta}{D_4} \, \Delta \, - \, 3 \frac{N^\Pi}{D_4} \, \Pi \Big)^\bullet
		+ C_5 h_{\theta\gamma} \dot{t}~^{<\theta\gamma >_3} +\frac{1}{c^2} \, \Big( 2\frac{N_3}{D_3} \,  -  \frac{1}{5} \frac{N_{31}}{D_3} \Big) q_{\gamma} \, \dot{U}^\gamma \, - \\
		& \hspace{2cm} 
		\, \Big(- \, \frac{1}{3} \rho c^2 \theta_{1,2}  \, + \, \frac{1}{4 \, c^2} \, \frac{N^\Delta}{D_4} \, \Delta \, + \,
		\frac{N^\Pi}{D_4} \, \Pi \Big) \,
	  \partial_\alpha \,  U^\alpha  \, +  q^\mu  \partial_\mu \, \Big( \frac{N_{31}}{D_3} \Big)\,  \,  -\\
		& \hspace{2cm}
		 \, \frac{N_{31}}{D_3} \, h^\mu_\alpha \,  \partial_\mu \, q^\alpha - 2 C_5 \, t^{< \mu }_{\,\,\ \, \,\gamma >_3} \partial_\mu \, U^\gamma = \frac{3}{\tau} \, \Big( \frac{1}{4c^2} \, \frac{N^\Delta}{D_4} \, \Delta \, + \, \frac{N^\Pi}{D_4} \, \Pi \Big),
	\end{split}
\end{align}
and the spatial traceless part is:
\begin{align}\label{devia}
\begin{split}
& C_5 \, h_{\gamma < \delta} \, h_{\beta >_3 \theta} \dot{t}^{< \gamma \theta >_3}\, + \, t_{< \delta \beta >_3}\, \dot{C}_5 \, + \,
\frac{2}{c^2} \, \left( \frac{N_3}{D_3} \, + \, \frac{1}{5} \, \frac{N_{31}}{D_3} \right) q_{< \delta} \, \dot{U}_{\beta >_3} \, + \\
& + \, 2 \, \left(- \, \frac{1}{3} \, \rho \, c^2 \theta_{1,2} \, + \, \frac{1}{4 \, c^2} \, \frac{N^\Delta}{D_4} \, \Delta \, + \,
\frac{N^\pi}{D_4} \, \pi \right) \,
h_{\gamma  < \delta } \, h^\mu_{\,\, \beta >_3} \, \partial_\mu \, U^\gamma  \, + \\
& + \frac{2}{5} \, \left(    q_{< \delta} h^\mu_{\,\, \beta >_3} \right) \, \partial_\mu \, \left( \frac{N_{31}}{D_3} \right)
- \, \frac{2}{5} \, \frac{N_{31}}{D_3} \,
\left(
h_{\gamma < \delta} h^\mu_{\,\, \beta >_3} \, \partial_\mu \, q^\gamma \right) + \\
& + C_5 \,\left[ t_{< \delta \beta >_3} \,  \partial_\alpha
\,  U^\alpha  \,+ \, 2 \,  t^{< \mu \gamma >_3}  h_{\gamma < \beta} \,
h_{\delta >_3  \nu} \,  \partial_\mu \, U^\nu  \right] =  - \, \frac{1}{\tau}
\,  C_5 \, t_{< \delta \beta >_3} \, .
\end{split}
\end{align}
The system formed by the $15$ equations \eqref{derivmat}$_{1,2,3}$, \eqref{tracciap}, \eqref{devia} and \eqref{derivmat}$_{5,6}$ is a closed systems for the $15$ unknown $(\rho, U_\delta,T,\Pi,t_{<\alpha \beta>_3}, q_\delta, \Delta)$.

\section{Entropy density, Convexity, Entropy Principle and well-posedness of Cauchy problem}\label{entro}
In this section we evaluate the entropy law and we want to prove that every solutions are entropic with an entropy density  that is a  convex function.
\subsection{Entropy density}
By substituting the distribution function \eqref{fgenE} with \eqref{RT} into \eqref{entropy}, we can evaluate the four-dimensional entropy flux. In this procedure, it is needed to be careful concerning the order of the nonequilibrium variables. The present linear constitutive equations is related to the entropy with the second order of the nonequilibrium variables. By taking into account up to the second order in the expansion of the distribution function and of the constitutive equations, we may evaluate as follows
\begin{align}\label{halpha}
 h^\alpha 
=  h^\alpha_E \, + \,  h^\alpha_{(1)} \, + \,  h^\alpha_{(2)}  \, ,
\end{align}
where $h^\alpha_{(1)}$ and $h^\alpha_{(2)}$ are, respectively, the contribution of the first and second order terms of the nonequilibrium variables, which can be derived as follows (see  Appendix~\ref{app:entropy} for details):
\begin{align}\label{entropyd1}
\begin{split} 
 h^\alpha_{(1)}  &= -\frac{c}{k_B}  \int_{\R^3} \int_{0}^{+ \infty} p^\alpha f_E \, \chi_E \, \tilde{\chi}_{(1)} \, \varphi ( \mathcal{I} ) \, d  \mathcal{I} \, d  \boldsymbol{P},   \\
 h^\alpha_{(2)}  &=  -  \frac{c}{2 \, k_B}  \int_{\R^3} \int_{0}^{+ \infty} p^\alpha f_E \, {\tilde{\chi}_{(1)}}{}^2  \, \varphi ( \mathcal{I} ) \, d \mathcal{I} \, d  \boldsymbol{P} \, ,
\end{split} 
\end{align}
where ${\tilde{\chi}_{(1)}}$ is $\tilde{\chi}$ defined in \eqref{fgenE} with the linear constitutive equations studied in the previous. After cumbersome calculations, we obtain explicit expression of them as follows:
\begin{align} \label{entropyd2}
 h^\alpha_{(1)} & =  \lambda_E \left(V^\alpha -  V^\alpha_E\right) \, + \, \frac{U_\mu}{T} \,  \left(T^{\alpha \mu} -  T^{\alpha \mu}_E\right) 
= \frac{q^\alpha}{T}, \nonumber \\
 h^\alpha_{(2)} &= -  \frac{m}{2 \, k_B} \left\{ \left[\left( \lambda-\lambda^E\right)\right]^2 V^\alpha_E \, + \, \left(\lambda_{\mu}-\lambda_{\mu}^E\right) \left(\lambda_{\nu}-\lambda_{\nu}^E\right) \, A^{\alpha \mu \nu}_E \, + \, \left(\lambda_{\mu \nu }\right) \left(\lambda_{\psi \theta }\right) \, A^{\alpha \mu \nu \psi \theta}_E \, +  \right. \nonumber \\
&\qquad  + \left.  2  \left( \lambda-\lambda^E\right) \left(\lambda_{\mu}-\lambda_{\mu}^E\right) \, T^{\alpha \mu}_E \, + \,  2  \left( \lambda-\lambda^E\right) \left(\lambda_{\mu \nu}\right) \, A^{\alpha \mu \nu}_E \, + \,  2  \left( \lambda_\theta-\lambda_\theta^E\right) \left(\lambda_{\mu \nu}\right) \, A^{\alpha \theta \mu \nu}_E \right\} \, \\
&= - \frac{1}{c^2}U^\alpha \left\{- \frac{c^2 \alpha_4 C_5}{2} t^{< \mu\nu >_3}t_{< \mu\nu>_3} -  \left(c^2 \alpha_3 \frac{N_3}{D_3} + \frac{b_3}{2}\right)q_\mu q^\mu 
+L_1\Pi^2 +L_2 \Delta^2 +2 L_3 \Pi \Delta \right\}\nonumber  \\
&\qquad + \frac{1}{2}\left(b_1-b_3 + c^2 \frac{N_3}{D_3}\alpha_1 + 
\frac{N_{31}}{D_3}\alpha_2 +2\alpha_3 c^2 \frac{N^\Pi}{D_4}\right)\Pi q^\alpha + \frac{1}{2}\left(b_2 + c^2 \frac{N_3}{D_3}\beta_1 + \frac{N_{31}}{D_3}\beta_2 +\frac{1}{2}\alpha_3 \frac{N^\Delta}{D_4}
\right)\Delta q^\alpha \nonumber \\
&\qquad + \frac{1}{2}\left(b_3 + 2c^2 \alpha_3 C_5 - \frac{2}{5}\alpha_4 \frac{N_{31}}{D_3}\right)t^{< \alpha\mu >_3}q_\mu, \nonumber 
\end{align}
where
\begin{align*}
 &L_1 = \frac{3c^2}{2}\alpha_2 \frac{N^\Pi}{D_4},\qquad
 L_2 = \frac{1}{8}\left(\ 3\beta_2\frac{N^\Delta}{D_4} -  c^2 \beta_1\right), \qquad
L_3 =\frac{1}{4} \left(\frac{3\alpha_2}{4}\frac{N^\Delta}{D_4}+3 c^2{\beta_2}\frac{N^\Pi}{D_4}- \frac{c^2\alpha_1}{4}
\right).
\end{align*}
In particular, for the entropy density $h=h^\alpha U_\alpha$, we have
\begin{align}\label{conv}
 h= h_E 
 + \frac{c^2 \alpha_4 C_5}{2} t^{< \mu\nu >_3}t_{< \mu\nu>_3} +  \left(c^2 \alpha_3 \frac{N_3}{D_3} + \frac{b_3}{2}\right)q_\mu q^\mu
-
 \begin{pmatrix}
  \Pi & \Delta
 \end{pmatrix}
\begin{pmatrix}
 L_1 & L_3\\
L_3 &  L_2
\end{pmatrix}
\begin{pmatrix}
 \Pi \\ \Delta
\end{pmatrix}.
\end{align}
We emphasize that the convexity of the entropy density is satisfied because from \eqref{entropyd2}$_1$ we have $h^\alpha_{(1)}U_\alpha =0$ and from   \eqref{entropyd1} we have $h^\alpha_{(2)}U_\alpha <0$ everywhere and zero only at equilibrium. Therefore the following inequalities are automatically satisfied:
\begin{itemize}
 \item $\alpha_4 C_5 <0$, 
 \item $\displaystyle 2c^2 \alpha_3 \frac{N_3}{D_3} + {b_3}>0$ because $q_\alpha q^\alpha <0$,
 \item $L_1>0$,
 \item $L_1 \, L_2 - \left(L_3\right)^2>0$.
\end{itemize}


\subsection{Entropy production}
According with the theorem proved by Boillat and Ruggeri \cite{BoillatRuggeri-1998} (see also \cite{RET,RS}), the procedure of MEP at molecular level it is equivalent to the closure using the entropy principle and the Lagrange multipliers coincide with the {\emph{main field}}
 for which the original system become symmetric hyperbolic \cite{RS}. Therefore the closed system satisfy the entropy balance law 
 \begin{equation}\label{ep}
\partial_\alpha h^\alpha = \Sigma,
\end{equation}
where   the entropy four-vector is given by \eqref{halpha}, \eqref{entropyd2}. For what concerns 
the entropy production $\Sigma$ according with the result of Ruggeri and Strumia \cite{RS} this  is given by  the scalar product between the  main field components  and the production terms \cite{RugStr}. In the present case, we have
\begin{align}\label{EP}
\Sigma = I^{\beta\gamma}\, \lambda_{\beta\gamma}.
\end{align}
By using eq. \eqref{P22} we have
\begin{align}\label{EP1}
\Sigma =\frac{1}{\tau}
\left\{    
- \frac{1}{4c^4  } \Delta  \,  U^\beta U^{\gamma } \lambda_{\beta\gamma} + \Big( \frac{1}{4c^2} \frac{N^\Delta}{D_4}\Delta +\frac{N^\Pi}{D_4} \Pi \Big) h^{\beta\gamma}\lambda_{\beta\gamma} + \Big( -\frac{2}{c^2 } \frac{N_3}{D_3} +\frac{\theta_{1,3}}{\theta_{1,2}}\frac{1}{c^2 } \Big)q^{(\beta} U^{ \gamma )}\lambda_{\beta\gamma} -   C_5 t^{<\beta\gamma>_3}\lambda_{\beta\gamma}
\right\}.\nonumber\\
\end{align}
By substituting eqs. \eqref{EP2}-\eqref{EP4} into eq. \eqref{EP1} and reminding that $q^\beta U^\gamma \lambda_{\beta\gamma}= - q_\alpha h^{\alpha \beta} U^\gamma \lambda_{\beta\gamma}$, we obtain $\Sigma$ in a quadratic form as follows:
\begin{align}\label{EP6}
 \Sigma = \frac{3 k_B \, C_5}{2 \tau m c^4 \rho\theta_{2,3}} t^{<\beta \gamma>_3} t_{<\beta \gamma>_3} + \frac{9 k_B \theta_{1,1}}{2 \tau m^2\, n\, c^6 D_3}\Big( -2 \frac{N_3}{D_3} +\frac{\theta_{1,3}}{\theta_{1,2}} \Big) q^\alpha q_\alpha
 + \begin{pmatrix}
	\Delta & \Pi
   \end{pmatrix}
 	\left(
\begin{matrix} 
		 M_1 & M_2  \\
		 M_2 & M_3
	\end{matrix} \right)
 \begin{pmatrix}
  \Delta \\ \Pi
 \end{pmatrix},
\end{align}
where 
\begin{align*}
\begin{split}
 &M_1 = \frac{k_B}{16\, c^8 \tau m^2 \,n\, D_4}\left(D_4^{33} + \frac{N^\Delta}{D_4}D_4^{34}\right), \qquad 
M_2 = - \frac{k_B}{4\, c^6 \tau m^2 \,n\, D_4} \left(D_4^{43}+\frac{N^\Delta}{D_4}D_4^{44} - \frac{N^\Pi}{D_4} D_4^{34}\right),\\
&M_3 = - \frac{k_B}{ c^4 \tau m^2 \,n\, D_4} \frac{N^\Pi}{D_4} D_4^{44}.
\end{split}
\end{align*}
The Sylvester criteria allows us to state that the quadratic form is positive definite iff all the following conditions hold:
\begin{enumerate}
	\item $\frac{3 k_B \, C_5}{2 \tau m c^4 \rho\theta_{2,3}}>0$.
	\item $\left( -2 \frac{N_3}{D_3} +\frac{\theta_{1,3}}{\theta_{1,2}} \right)\frac{9 k_B \theta_{1,1}}{2 \tau m^2\, n\, c^6 D_3} <0$, because $q^\alpha q_\alpha <0$.
	\item $M_1 >0$.
	\item $M_1 \,  M_3 - (M_2)^2 >0$.
\end{enumerate}
The first condition is automatically satisfied because of the definition of the functions involved.\\
In order to prove the second condition, we can consider a space like vector $X^\beta$ and the following function that is defined to be positive for each value of $X^\beta$.
\begin{align*}
	g(X^\beta) = \frac{U_\alpha}{c \,\tau \, k_B} \int_{\R^3} \int_0^{+\infty} f_E p^\alpha \left[ X_\beta p^\beta \left( \frac{\theta_{1,3}}{\theta_{1,2}} \left( 1+ \frac{\mathcal{I}}{mc^2}\right) - \frac{2}{mc^2} \left( 1+ \frac{\mathcal{I}}{mc^2}\right)^2 U_\nu p^\nu\right)\right]^2 \phi(\mathcal{I}) d \mathcal{I} d\boldsymbol{P} \, .
\end{align*}
By exploiting the calculation in the above integral and by using eqs. \eqref{A1w}, we have
\begin{align*}
	g(X^\beta) = \frac{m^2 \, n \, c^2}{\tau \, k_B} \left[ \frac{1}{3}\frac{\left(\theta_{1,3}\right)^2}{\theta_{1,2}} - \frac{2}{5} \theta_{1,4}\right] X^\beta \, X_\beta \, .
\end{align*}
If we choose, as a particular value,
\begin{align*}
	X^\beta = - \frac{1}{D_3} \frac{9 \, k_B}{2 \, m^2 \, n \, c^4} \theta_{1,1}\, q^\beta ,
\end{align*}
we obtain
\begin{align*}
	g(X^\beta) = \frac{9 k_B \theta_{1,1}}{2 \tau m^2\, n\, c^6 D_3}\Big( -2 \frac{N_3}{D_3} +\frac{\theta_{1,3}}{\theta_{1,2}} \Big) q^\alpha q_\alpha > 0.
\end{align*}
It proves that also the second condition is satisfied.\\
Conditions 3 and 4 can be proved showing that they are coefficients of a quadratic form that is definite positive. In order to obtain the entropy production up to the second order, we have to substitute eq. \eqref{relRETpol}$_4$ into \eqref{EP} and take the collisional term \eqref{P1} up to the first order. Then, 
\begin{align*}
\Sigma^{(2)} = \frac{c}{m}\int_{\R^{3}}
\int_0^{+\infty} Q^{(1)}  \,  p^{\beta}  p^{\gamma}\,\lambda_{\beta\gamma} \, \Big(1 +  \frac{\II}{mc^2} \Big)^2\, 
\phi(\mathcal{I}) \, d \mathcal{I} \, d \boldsymbol{P},
\end{align*}
with \begin{align*}
	Q^{(1)} = \frac{f_E}{c^2 \, \tau \, k_B} U_\alpha p^\alpha \left[ \tilde{\chi} - \frac{k_B}{bmc^2} p^\mu q_\mu \left(1+\frac{\mathcal{I}}{mc^2}\right)\right].
\end{align*}
If we substitute to $\lambda_{\beta \gamma }$ its expression obtained from eq. \eqref{fgenE}$_2$, we obtain
\begin{align*}
	\Sigma^{(2)} = {c}\int_{\R^{3}}
	\int_0^{+\infty} Q^{(1)}  \, \tilde{\chi} \, \phi(\mathcal{I}) \, d \mathcal{I} \, d \boldsymbol{P}.
\end{align*}
In the state where $q^\beta =0$ and $t^{<\alpha \beta >_3}=0$ the Lagrange multipliers and the Entropy production assume particular values that we denote with a $*$, in particular
\begin{align*}
	\Sigma^{(2*)} = \frac{c}{m}\int_{\R^{3}}
	\int_0^{+\infty} Q^{(1*)}  \, \tilde{\chi}^* \, \phi(\mathcal{I}) \, d \mathcal{I} \, d \boldsymbol{P} = \frac{c}{m}\int_{\R^{3}}
	\int_0^{+\infty}\frac{f_E}{c^2 \, \tau \, k_B} U_\alpha p^\alpha \, \left[\tilde{\chi*}\right]^2 \phi(\mathcal{I}) \, d \mathcal{I} \, d \boldsymbol{P},
\end{align*}
that is clearly a positive quantity. Moreover, we have 
\begin{align*}
	\Sigma^{(2*)} = I^{\beta\gamma *}\lambda_{\beta \gamma *}
\end{align*}
that corresponds to the quadratic form 
\begin{align*}
\begin{pmatrix}
	\Delta & \Pi
\end{pmatrix}
\left(
\begin{matrix} 
	M_1 & M_2  \\
	M_2 & M_3
\end{matrix} \right)
\begin{pmatrix}
	\Delta \\ \Pi
\end{pmatrix},
\end{align*}
which, therefore, turns to be definite positive.
Therefore proved the 
\begin{statement}
The entropy density  \eqref{conv}  is a  convex function and is maximum at equilibrium. The solutions satisfies the entropy principle \eqref{ep} with an entropy production \eqref{EP6} that is always non negative. 
According with the general theory of symmetrization given first in covariant formulation in \cite{RugStr},  and the equivalence between Lagrange multipliers and main field \cite{BoillatRuggeri-1998},   the closed system is symmetric hyperbolic in the neighborhood of equilibrium if we chose as variables the main field variables \eqref{RT}, with coefficients given in \eqref{coef1}, \eqref{coef2}, \eqref{coef3} and the Cauchy problem is well posed locally in time.
\end{statement}

\section{Diatomic gases}\label{diatomico}
The system \eqref{derivmat} is very complex, in particular, because it is not simple to evaluate the function $\omega(\gamma)$ that involves two integrals \eqref{10}$_3$ that cannot have analytical expression for a generic polyatomic gas. 
Taking into account the relations \cite{Annals}
\begin{equation*}
J_{2,1} (\gamma) = \frac{1}{\gamma}K_2(\gamma), \qquad \qquad J_{2,2} (\gamma) = \frac{1}{\gamma}\Big(K_3(\gamma)-\frac{1}{\gamma}K_2(\gamma)\Big), 
\end{equation*}
where $K_n$ denotes the modified Bessel function, we can rewrite $\omega$ given in \eqref{10}$_3$ in terms of the modified Bessel functions \cite{Annals}:
\begin{equation*}
\omega(\gamma) = \frac{1}{\gamma}\left(
\frac{\int_0^{+\infty} K_3(\gamma^*)\, \phi(\mathcal{I})  \, d \, \mathcal{I}}
{\int_0^{+\infty}\frac{K_2(\gamma^*)}{\gamma^*}\phi(\mathcal{I})  \, d \, \mathcal{I}}-1\right).
\end{equation*}
Moreover, to calculate the integrals, we need to prescribe the measure $\phi(\mathcal{I})$.
In \cite{Annals}, the measure $\phi(\II)$ was assumed as
\begin{equation*}
\phi(\mathcal{I})  = \mathcal{I}^a, \qquad a = \frac{D-5}{2},
\end{equation*}
because it is the one for which the macroscopic internal energy in the classical limit, when $\gamma \rightarrow \infty$ , converges to the one of a classical polyatomic gas, where  $D$ indicates the degree of freedom of molecule. 
As was observed by Ruggeri, Xiao and Zhao  \cite{Xiao} in the case of $a=0$, i.e., $D=5$ corresponding to diatomic gas, the energy $e$  have an explicit expression similar to the monatomic gas:
\begin{equation*}
e = p \left(\frac{\gamma K_0(\gamma)}{K_1(\gamma)}+3\right).
\end{equation*} 
Therefore, from \eqref{10}, we have
\begin{equation*}\label{omegadiat}
\omega_{\text{diat}}(\gamma)= \frac{K_0(\gamma)}{K_1(\gamma)} +\frac{3}{\gamma}.
\end{equation*}
Using the following recurrence formulas of the Bessel functions
\begin{align}\label{recur}
K_n(\gamma) = \frac{\gamma}{2n} \Big(K_{n+1}(\gamma) - K_{n-1}(\gamma) \Big) \, ,
\end{align}
we can express $\omega$ in terms of
\begin{equation*}
G(\gamma) = \frac{K_3(\gamma)}{K_2(\gamma)}.
\end{equation*}
In fact we can obtain immediately the following expression:
\begin{align}\label{omegaD}
\omega_{\text{diat}}(\gamma)=     \frac{1}{\gamma } + \frac{ \gamma}{\gamma G -4},
\end{align}
which is a simple function similar to the one of monatomic gas for which we have \cite{LMR}:
\begin{align*}
    \omega_{\text{mono}}(\gamma) =   -1 +  \, \gamma \, G.
\end{align*}
Taking into account that the derivative of Bessel function are known, all coefficients appearing in the differential system \eqref{derivmat} can be written explicitly in terms of $G(\gamma)$, by using \eqref{omegaD} and the recurrence formula \eqref{recur}. This is simple by using a symbolic calculus like Mathematica\textregistered.


\section{Ultra-relativistic limit}\label{sec:ultra}
In the ultra-relativistic limit where $\gamma \rightarrow 0$, it was proved in  \cite{5} and \cite{6} that the energy converges to 
\begin{align}\label{25}
    e = (\alpha +1) \frac{n \, m \, c^2}{\gamma} \, , \quad \mbox{with} \quad \alpha= \left\{
    \begin{matrix}
        2 & \mbox{if} \quad a \leq 2 \\
        a & \, \, \,  \mbox{if}  \quad a \geq 2 \, .
    \end{matrix} \right.
\end{align}
This implies
\begin{align}\label{omefaultre}
    \omega_{\text{ultra}} = \frac{(\alpha +1) }{\gamma} \, , \quad \mbox{with} \quad \alpha= \left\{
    \begin{matrix}
        2 & \mbox{if} \quad a \leq 2 \\
        a & \, \, \,  \mbox{if}  \quad a \geq 2 \, .
    \end{matrix} \right.
\end{align}
By means of this expression, we can evaluate the coefficients $\theta_{h,j}$ in \eqref{thetas} that become:
\begin{align*} 
\begin{split}
& \left\{\theta_{0,0},\theta_{0,1},\theta_{0,2},\theta_{0,3},\theta_{0,4}\right\} = \\
& \hspace{1cm }\left\{1,\frac{\alpha
   +1}{\gamma },\frac{(\alpha
   +1) (\alpha +2)}{\gamma
   ^2},\frac{(\alpha +1)
   (\alpha +2) (\alpha
   +3)}{\gamma
   ^3},\frac{(\alpha +1)
   (\alpha +2) (\alpha +3)
   (\alpha +4)}{\gamma
   ^4}\right\}, \\
 & \left\{\theta_{1,1},\theta_{1,2},\theta_{1,3},\theta_{1,4}\right\} = \\
 & \hspace{1cm } 
  \left\{\frac{1}{\gamma
      },\frac{3 (\alpha
      +2)}{\gamma ^2},\frac{6
      (\alpha +2) (\alpha
      +3)}{\gamma ^3},\frac{10
      (\alpha +2) (\alpha +3)
      (\alpha +4)}{\gamma
      ^4}\right\}, \\
      &  \left\{\theta_{2,3},\theta_{2,4}\right\} = \\
       & \hspace{1cm } 
      \left\{\frac{3 (\alpha
         +2)}{\gamma ^3},\frac{15
         (\alpha +2) (\alpha
         +4)}{\gamma ^4}\right\}.
\end{split}
\end{align*}
It follows that, in the ultra-relativistic limit, we have
\begin{align*}
    \frac{N_3}{D_3} =  \frac{2(\alpha +3)}{\gamma} \quad, \quad \frac{N_{31}}{D_3} =  \frac{10}{\gamma}  \quad, \quad C_5 = \frac{\alpha +4}{\gamma},
\end{align*}
and
\begin{align}\label{26}
    \frac{N^\Pi}{D_4} = - \, \frac{\alpha +4}{\gamma} \quad, \quad \frac{N^\Delta}{D_4} = - \, \frac{1}{\alpha +1} \, ,
\end{align}
where the last two equations holds for $\alpha \neq 2$ (i.e., $a \neq 2$). For $a = 2$, the ultra-relativistic limit of $\frac{N^\Pi}{D_4}$ and of $\frac{N^\Delta}{D_4}$ gives the indeterminate form $\left[\frac{0}{0}\right]$.  We show (see Appendix~\ref{app:cont} for details) that it can be solved by considering higher order terms for the energy $e$, allowing to prove that eqs. \eqref{26} are valid also with $a=2$ and, hence, that the closure of the present model is continuous with respect to the parameter $\alpha$, at the ultra-relativistic limit.

\section{Principal subsystems of ET$_{15}$} \label{sec:sub}


For a general system of the hyperbolic system of balance laws, the system with smaller set of the field equations can be deduced retaining the property that the convexity of the entropy and the positivity of the entropy production  is preserved as a principal subsystem \cite{BoillatRuggeriARMA}. The principal subsystem is obtained by putting some components of the main field as a constant, and the corresponding balance laws are deleted. 

Let us recall the system \eqref{Annalis}. The balance law of $A^{\alpha \beta \gamma}$ is divided into the trace part $A^{\alpha \beta}_\beta$ and the traceless part $A^{\alpha < \beta\gamma>}$. As we study below, by deleting the trace part and putting the corresponding component of main field as zero, we obtain the theory with $14$ fields (ET$_{14}$). On the other hand, by conducting the same procedure on the traceless part, we obtain the theory with $6$ fields (ET$_{6}$). It is remarkable that ET$_{14}$ and ET$_{6}$ is the same order in the sense of the principle subsystem differently from the classical case in which the classical ET$_{6}$ is a principal subsystem of classical ET$_{14}$. Moreover, the relativistic Euler theory is deduced as a principal subsystem by deleting the balance laws of $A^{\alpha\beta\gamma}$ and putting the corresponding component of main field as zero

\subsection{ET$_{14}$: 14 fields theory}
The ET$_{14}$ is obtained as a principal subsystem of ET$_{15}$ under the condition $\lambda^\alpha_\alpha = 0$. From \eqref{RT}$_3$, this condition provides $\Delta$ expressed by $\Pi$ as follows:
\begin{align}
 \Delta^{(14)} &= - \frac{c^2\alpha_1 -{3}\alpha_2}{c^2 \beta_1 - {3}\beta_2}\Pi 
= 4 \frac{N_a}{D_a}c^2 \Pi, \label{subd}
\end{align}
where $N_a = D_4^{44}+D_4^{43}$ and  $D_a = D_4^{34}+D_4^{33}$.  Then, the independent fields are the following $14$ fields; $(\rho, \gamma, \Pi, U^\alpha, q^\alpha, t^{< \alpha\beta >_3})$. 
By deleting the balance equation corresponding to $\lambda^\alpha_\alpha$, that is, the one of $A^{\alpha \beta}_\beta$, the present system of the balance equations are the following:
\begin{align}\label{Sub14}
	\partial_\alpha V^\alpha =0 \, , \quad  \partial_\alpha T^{\alpha \beta} =0 \, , \quad   \partial_\alpha A^{ \alpha < \beta \gamma>} = I^{<\beta \gamma>} .
\end{align}

With \eqref{subd}, the constitutive equation is modified in this subsystem. 
For the comparison with the ET$_{14}$ theory studied in \cite{Annals}, let us denote
\begin{align*}
 \frac{N_1^\pi}{D_1^\pi}  = -\frac{1}{3}\frac{N_a}{D_a} , \quad
 \frac{N_{11}^\pi}{D_1^\pi} = \frac{1}{D_4}\left(\frac{N_a}{D_a}N^\Delta + N^\Pi \right).
\end{align*}
We can prove the following identity:
\begin{align*}
 \frac{N_b}{D_a}  = - \frac{1}{D_4}\left(\frac{N_a}{D_a}N^\Delta + N^\Pi \right), \quad \text{with} \quad N_b = N^{\Delta 34}+N^{\Delta 33},
\end{align*}
where $N^{\Delta 33}$ and $N^{\Delta 34}$ are the minor determinants of  $N^\Delta$ which deletes the third row and third column, and the third row and fourth column, respectively.
Then, as a result, instead of \eqref{24}, the closure for $A^{\alpha \beta \gamma}$ in the present principal subsystem is given by 
\begin{align}\label{Sub2}
	A^{\alpha \beta \gamma} = \left( \rho \, \theta_{0,2} - \frac{3}{c^2} \, \frac{N^\pi_1}{D_1^\pi} \, \Pi \right) U^\alpha U^\beta U^\gamma \, + \,  \left( \rho \, c^2 \theta_{1,2} - 3 \, \frac{N_{11}^\pi}{D_1^\pi} \, \Pi \right) U^{( \alpha} h^{\beta \gamma )} \, + \\
	+ \, \frac{3}{c^2} \, \frac{N_3}{D_3} \, q^{( \alpha} U^\beta U^{\gamma )}  \, + \, \frac{3}{5} \, \frac{N_{31}}{D_3} \, q^{( \alpha} h^{\beta \gamma )}  \, + \, 3 \, C_5 \, t^{( < \alpha \beta >_3} U^{\gamma )} \, . \nonumber
\end{align}
This result is formally same as the result of \cite{Annals} (eq. (56) of the paper). However, there are the difference in the coefficients due to the presence of $\left( mc^2 + \mathcal{I} \right)^n$ instead of $ mc^2 + \, n \, \mathcal{I}$ in the integrals.

Similarly, we obtain the production term in this principal subsystem as follows:
\begin{align}\label{Sub3}
	I^{< \beta \gamma >} = - \, \frac{1}{c^2 \tau} \,\frac{3N_1^\pi+N_{11}^\pi}{D_1^\pi} \, \Pi \, U^{< \beta}  U^{\gamma >}  \, + \, \frac{1}{c^2 \tau}   \left( \frac{\theta_{1,3}}{\theta_{1,2}}  \, - \, 
	2 \,  \frac{N_3}{D_3} \right) \, q^{( \beta} U^{\gamma )} \, - \, \frac{1}{\tau} \, C_5 \, t^{< \beta \gamma >_3} \, .
\end{align}
This expression \eqref{Sub3} is formally the same with the result of \cite{Car1} (eq (16) of the paper) in \cite{Car1}, except that now we have $\frac{\theta_{1,3}}{\theta_{1,2}} $ instead of $\frac{B_2}{B_4}$ defined in \cite{Car1} and the difference of the integral in the coefficients is similar with the case for $A^{\alpha\beta\gamma}$.

\emph{The system} \eqref{Sub14}   \emph{is symmetric hyperbolic in the main field} $(\lambda,\lambda_{\alpha}, \lambda_{< \mu \nu >})$ \emph{given respectively by} \eqref{RT} \emph{with} $\Delta = \Delta^{(14)}$ \emph{given by} \eqref{subd}.

\subsection{ET$_6$: 6 fields theory}

We consider the principal subsystem with $\lambda_{< \mu \nu >} = \lambda_{\mu \nu} - \frac{1}{4}\lambda_\alpha^\alpha g_{\mu \nu} = 0$, then we have
\begin{align} \label{lag6}
 \lambda_{\mu \nu} = \frac{1}{4}\lambda^\alpha_\alpha g_{\mu \nu}.
\end{align}
By comparing it with \eqref{RT}, we have
\begin{align*}
 \left(\alpha_1 + \frac{\alpha_2}{c^2}\right)\Pi + \left(\beta_1 + \frac{\beta_2}{c^2}\right)\Delta =0, \qquad q_\mu = 0, \qquad t_{< \mu\nu >_3} =0.
\end{align*}
The first equation indicates that, in this principal subsystem, $\Delta$ is expressed with $\Pi$ as follows:
\begin{align}\label{subd6}
 \Delta^{(6)} = - \frac{c^2 \alpha_1 + {\alpha_2}}{c^2 \beta_1 + {\beta_2}}\Pi = w \, \Pi
\end{align}
where 
\begin{equation*}
w = 4 c^2 \frac{D^{44}_4 - 3D^{43}_4 }{D^{34}_4 - 3D^{33} _4}.
\end{equation*}
It should be mentioned that the relation between $\Delta$ and $\Pi$ is different from the case of ET$_{14}$.

The independent fields are now the $6$ fields; $(n, \gamma, U^\alpha, \Pi)$, and the balance equations are the following:
\begin{align}\label{Sub61}
	\partial_\alpha V^\alpha =0 \, , \quad  \partial_\alpha T^{\alpha \beta} =0 \, , \quad   \partial_\alpha A^{\alpha \beta}_{\,\,\, \, \,\beta} = I^\beta_{\,\,\beta} \, .
\end{align}
where the energy-momentum tensor is now given, instead of \eqref{19}, by
\begin{align}\label{T6}
 T^{\alpha \beta} = \frac{e}{c^2} \,  U^{\alpha } U^\beta + \, \Big(p \, + \, \Pi\Big)
    h^{\alpha \beta}. 
\end{align}
and, from \eqref{24},
\begin{align}\label{Aa6}
 A^{\alpha \beta}_{\,\,\, \, \,\beta}  = \left\{\rho c^2 (\theta_{0,2} - \theta_{1,2}) + A_1\right\}\Pi U^\alpha,
\end{align}
where
\begin{align*}
 A_1 = - \frac{1}{4c^2}\left\{\left(1+3\frac{N^\Delta}{D_4}\right)\frac{c^2 \alpha_1 + {\alpha_2}}{c^2 \beta_1 + {\beta_2}} - 12c^2 \frac{N^\Pi}{D_4}\right\} = \frac{D_4^{44}-3D_4^{43} + 3 N^{\Delta 34} - 9N^{\Delta 33}}{D_4^{34} - 3 D_4^{33}}.
\end{align*}
Similarly, from \eqref{P22}, we obtain 
\begin{align}
I^\beta_{\,\,\beta} = - \frac{A_1}{\tau}\Pi.
\end{align}
 The corresponding Lagrange multiplier to $A^{\alpha \beta}_{\,\,\, \, \,\beta}$ is $\mu = \frac{1}{4}\lambda^\alpha_\alpha$ which is obtained from \eqref{lag6} as follows:
\begin{align}\label{mumu}
 \mu = c^2 \frac{\alpha_1\beta_2 - \alpha_2 \beta_1}{c^2 \beta_1 + \beta_2}\Pi.
\end{align}
\emph{The system} \eqref{Sub61} \emph{with} \eqref{T6} \emph{and} \eqref{Aa6} \emph{is symmetric hyperbolic in the main field} $(\lambda,\lambda_{\alpha}, \mu)$ \emph{given respectively by  (see} \eqref{RT}$_{1,2}$ \emph{):}
\begin{equation}
\lambda = -\frac{g+c^2}{T} +(a1+ a2 w)\,\Pi, \qquad \lambda_\alpha = \frac{1}{T}\left(1+(b1+b2 w) \Pi
\right) U_\alpha,
\end{equation}
\emph{and $\mu$  given by }\eqref{mumu}.

The closed field equations with the material derivative are obtained as follows:
\begin{align}\label{field6}
	\begin{split}
		& \dot{\rho}  + \rho \, \partial_\alpha \, U_\alpha=0 \, , \, \\
&\frac{e+p+ \Pi}{c^2} \, \dot{U}_\delta   \, + \,  h_\delta^\mu \, \partial_\mu (p+ \Pi) =0 \, , \, \\
&\dot{e} \, + \,  (e+p+ \Pi)\,  \partial_\alpha U^\alpha =0 \, , \\
& \dot{\Pi} \, + \, \frac{\rho c^2 \left({\theta}_{0,2}' -  {\theta}_{1,2}' \right)}{A_1}\dot{\gamma}  \, + \, \frac{\dot{A}_1}{A_1} \Pi    \, + \,   \Pi \partial_\alpha U^\alpha  =  - \, \frac{\Pi}{\tau} .
	\end{split}
\end{align}
Taking into account
\begin{equation}
 h_\delta^\mu \, \partial_\mu (p+ \Pi)  = U_\delta \frac{\dot{p}+\dot{\Pi}}{c^2} - \partial_\delta (p+\Pi),
\end{equation}
and from \eqref{10}:
\begin{equation}
\dot{e} = c^2( \dot{\rho} \omega + \rho \omega^\prime \dot{\gamma}), \qquad \dot{p}=\frac{c^2}{\gamma^2}(\gamma \dot{\rho} - \rho \dot{\gamma}),
\end{equation}
the system \eqref{field6} can be put in the normal form:
\begin{align}\label{field6bis}
		& \dot{\rho}  + \rho \, \partial_\alpha \, U_\alpha=0 \, , \, \nonumber\\
&\left(\rho +\frac{\rho \varepsilon+p+ \Pi}{c^2}\right) \, \dot{U}_\delta   \, - \partial_\delta (p+ \Pi) 
-  \frac{(p+\Pi)}{c^2}\left[1- \frac{1}{A_1 \omega^\prime}\left(
A_1^\prime \frac{\Pi}{\rho c^2} + \frac{A_1}{\gamma^2} +\theta_{0,2}^\prime-\theta_{1,2}^\prime
\right) 
\right] U_\delta \, \partial_\alpha U^\alpha
= \frac{\Pi}{\tau c^2 }U_\delta
 \, , \, \nonumber\\
&\rho c^2 \omega^\prime\, \dot{\gamma} \, + \,  (p+ \Pi)\,  \partial_\alpha U^\alpha =0 \, , \\
& \dot{\Pi} \, + \left\{\Pi -\frac{p+\Pi}{\rho c^2 A_1 \omega^\prime}\left[A_1^\prime \Pi +\rho c^2\left({\theta}_{0,2}' -  {\theta}_{1,2}' \right)\right]\right\}\partial_\alpha U^\alpha 
   =  - \, \frac{\Pi}{\tau} .\nonumber
\end{align}
It is extremely interesting that in the relativistic theory the acceleration is influenced by the relaxation  time trough the right hand side of \eqref{field6bis}$_2$ and this  may be important for the application to the problems of cosmology.

\subsection{ET$_5$:  Euler 5 fields theory}
Let us consider the principal subsystem with $\lambda_{\mu \nu}=0$. This indicates that any nonequilibrium variables are set to be zero, i.e.,
\begin{align}
 \Pi = \Delta=0, \qquad t_{< \mu\nu>_3} =0, \qquad q_\alpha=0.
\end{align}
The independent fields are the $5$ fields $(n, U^\alpha, \gamma)$, and the balance equations are
\begin{align}\label{Eulers}
	\partial_\alpha V^\alpha =0 \, , \qquad   \partial_\alpha T^{\alpha \beta} =0,
\end{align}
with
\begin{align}
 T^{\alpha \beta} = \frac{e}{c^2} \,  U^{\alpha } U^\beta + \, p h^{\alpha \beta}. 
\end{align}
\emph{The deduced system is the one of the relativistic Euler theory and the system \eqref{Eulers} become symmetric in the main field $(\lambda = -(g+c^2)/T,\lambda_\alpha = U_\alpha/T)$ obtained first by  Ruggeri and Strumia in \cite{RugStr}.}

\section{Maxwellian Iteration and phenomenological coefficients}\label{sec:max}
In order to find the parabolic limit of system \eqref{derivmat} and to obtain the corresponding Eckart equations, we adopt the Maxwellian iteration \cite{Ikenberry} on \eqref{derivmat} in which only the first order terms with respect to the relaxation time are retained. The phenomenological coefficients, that is, the heat conductivity $\chi$, the shear viscosity $\mu$ and the bulk viscosity $\nu$ are identified with the relaxation time.

The method of the Maxwellian iteration is based on putting to zero the nonequilibrium variables on  the left side of equations \eqref{derivmat}:
\begin{align}\label{Pen3}
\begin{split}
& \dot{\rho}  - \rho \, h^{\beta \alpha} \, \partial_\beta \, U_\alpha=0 \, , \\
& \frac{e+p}{c^2}\,  h_{\delta \beta} \, \dot{U}^\beta  \, - \, h_\delta^\mu \, \partial_\mu p  =0 \, , \\
& \dot{e} \,  - \, (e+p)\,  h_\alpha^\mu  \, \partial_\mu U^\alpha  =0 \, , \\
& \frac{c^2}{3} h_{\delta \beta} \, \Big(\dot{\rho} \theta_{1,2}+\rho  \theta'_{1,2}\dot{\gamma}\Big)
 - \,  \frac{1}{3} \rho c^2 \theta_{1,2}   \,
\left[ h_{\delta \beta}  \, h_\alpha^\mu  \, \partial_\mu \,  U^\alpha \, + \, 2
\, h_{\theta ( \delta } \, h_{\beta )}^\mu \, \partial_\mu \, U^\theta \right] = \\
& \quad  = \frac{1}{\tau} \, \Big( \frac{1}{4} \, \frac{N^\Delta}{D_4} \,
\frac{\Delta}{c^2} \, + \, \frac{N^\Pi}{D_4} \, \Pi  \Big) h_{\delta \beta} \, - \, \frac{1}{\tau}
\,  C_5 \, t_{< \delta \beta >_3}\, , \\
& h_{\beta \delta} \, \dot{U}^\beta \Big( \rho  \theta_{0,2}  c^2 + 2 \frac{1}{3} \rho c^2 \theta_{1,2} \Big)  \, - \,
 h_\delta^\mu \, c^2 \, \partial_\mu \, \frac{1}{3} \rho c^2 \theta_{1,2}    =
\frac{1}{\tau} \,  \Big( \frac{N_3}{D_3} \, - \, \frac{\theta_{1,3}}{2 \, \theta_{1,2}} \Big) \, q_\delta  \, , \\
&c^4 \Big(\dot{\rho} \theta_{0,2} +  \rho \theta'_{0,2}\dot{\gamma} \Big)  \, - \,
 \, \rho c^4 \Big(  \theta_{0,2} +  \frac{2}{3} \theta_{1,2}   \Big)  h_\alpha^\mu \, \partial_\mu \, U^\alpha \,  =  - \, \frac{1}{4 \, \tau} \, \Delta  \, .
\end{split}
\end{align}
From the first three  equations of \eqref{Pen3} and taking into account $p=\rho c^2/\gamma, e = \rho c^2 \omega(\gamma) $ (see \eqref{10}), we can deduce
\begin{align}\label{3con}
\dot{\rho}  = \rho \, h^{\mu \alpha} \partial_\mu \, U_\alpha , \quad 
h_\delta^\mu \, \partial_\mu \, \rho = \rho \frac{\omega \gamma +1}{ c^2} \, h_{\delta \beta} U^\mu \, \partial_\mu \, U^\beta \, + \, \frac{\rho}{\gamma} \,
  h_\delta^\mu \, \partial_\mu \, \gamma, \quad  \, 
 \dot{\gamma} = \frac{1}{\gamma \omega'} \, h^{\mu \alpha} \partial_\mu \, U_\alpha \, .
\end{align}
Putting \eqref{3con} in the remaining equations \eqref{Pen3}$_{4,5,6}$, we obtain the solution
\begin{align}\label{I8b}
    \begin{split}
 &   q_{\beta}=-\chi \,  h_\beta ^{\alpha}\left [  \partial_{\alpha}T- \frac{T}{c^2}U^{\mu}\partial_{\mu}U^{\alpha}\right ], \\
  & \Pi= -\nu \, \partial_{\alpha} U^{\alpha}, \\
       &   t_{<\beta \delta>_3} =2 \mu  \, h^{\alpha}_{\beta}\, h^{\mu}_{\delta}\partial_{<\alpha} U_{\mu>}, \\
        &   \Delta = \sigma \, \partial_{\alpha} U^{\alpha},
    \end{split}
\end{align}
with
\begin{align}\label{I8n}
    \begin{split}
    & \chi = - \frac{2\rho c^2}{3 B^q T } \left[3 \theta_{0,2} + \theta_{1,2} (1-\omega \,\gamma)\right],\\
        & \nu= -\frac{\rho c^2}{3B_2^\Pi} \left\{ 
         \frac{2}{3} \theta_{1,2} - \frac{\theta'_{1,2}}{\gamma \omega' }+ 3 \frac{N^\Delta}{D_4} \Big(\frac{2}{3} \theta_{1,2} - \frac{\theta'_{0,2}}{\gamma \omega'}
         \Big)
        \right\}, \\
        & \mu = -\frac{\rho c^2}{3B^t}  \theta_{1,2}\, ,
         \end{split}
\end{align}
and
\[
  \sigma = \frac{\rho }{B_1^\Delta}  \Big(  \frac{2}{3} \theta_{1,2} - \frac{\theta'_{0,2}}{\gamma \omega'} \Big),
\]
where $B_2^\Pi, \, B^q,\, B^t$ are explicitly given by \eqref{P3} with the relaxation time $\tau$.

As the first three equations in \eqref{I8b} are the Eckart equations, we deduce that $\chi, \nu, \mu$ are respectively  the heat conductivity, the bulk and shear viscosities. In addition, we have a new phenomenological coefficient $\sigma$ but as $\Delta$ doesn't appears neither in $V^\alpha$ and $T^{\alpha\beta}$ (see the equation \eqref{19} or the first three equations in \eqref{derivmat}), we arrive to the conclusion that the present theory converges to the Eckart one formed in the first three block equations  of \eqref{derivmat} with constitutive equations \eqref{I8b} in which the heat conductivity, bulk viscosity and  shear viscosity are explicitly given by \eqref{I8n}$_{1,2,3}$. 

We introduce as in \cite{Car2} the dimensionless variables as follows:
\begin{align}\label{I8nn}
    \begin{split}
    &\bar{\chi} = \frac{\rho T \chi}{p^2 \tau} = -\frac{2}{3} \gamma^2   \frac{3 \theta_{0,2} + \theta_{1,2} (1-\omega \,\gamma)}{\frac{\theta_{1,3}}{\theta_{1,2}} \, - 2 \frac{N_3}{D_3} },\\
        & \bar{\nu} = \frac{\nu}{p \tau}= -\frac{1}{3} \frac{\gamma}{\frac{N^\Pi}{D_4}}   \left\{ 
                 \frac{2}{3} \theta_{1,2} - \frac{\theta'_{1,2}}{\gamma \omega' }+ 3 \frac{N^\Delta}{D_4} \Big(\frac{2}{3} \theta_{1,2} - \frac{\theta'_{0,2}}{\gamma \omega'}
                 \Big)
                \right\}, \\
        & \bar{\mu} = \frac{\mu}{p \tau}= \frac{\gamma}{3C_5}  \theta_{1,2}\, ,
         \end{split}
\end{align}
that are functions only of $\gamma$.  
\subsection{Ultra-relativistic and classical limit of the phenomenological coefficients}
Taking into account eqs. \eqref{omefaultre} - \eqref{26}, it is simple to obtain the limit of \eqref{I8nn} when $\gamma \rightarrow 0$:
 \begin{align*}
     \begin{split}
     {\bar{\chi}}_\text{ultra} = 0,\qquad
          {\bar{\nu}} _\text{ultra} = \frac{2}{3} \frac{\alpha^2 -4}{(1 + \alpha) (4 + \alpha)}, \qquad
          {\bar{\mu} }_\text{ultra} =\frac{2 + \alpha}{4 + \alpha}.
          \end{split}
 \end{align*}
 In particular in the most significant case in which $a\leq 2$ for which $\alpha =2$ we have
  \begin{align}\label{I8ultran}
      \begin{split}
      {\bar{\chi}}_\text{ultra} = 0,\qquad
           {\bar{\nu}} _\text{ultra} = 0, \qquad
           {\bar{\mu} }_\text{ultra} =\frac{2}{3}.
           \end{split}
  \end{align}
  
Instead, in the classical limit for which $\gamma \rightarrow \infty$, it was proved in \cite{Annals} that the internal energy $\varepsilon$ converges to the classical internal energy of polytropic gas: $\varepsilon = (D/2)(k_B/m)T$. Therefore, from \eqref{interenergy2}, $\omega$ converges to
\begin{equation}\label{wclas}
\omega_{\text{class}} = 1+ \frac{D}{2 \gamma}.
\end{equation}
 In the present case, using \eqref{wclas}, it is not difficult to find $\theta_{h,j}$ deduced in \eqref{thetas} in the limit $\gamma \rightarrow \infty$ as follows:
 \begin{align} \label{thetaclass}
 \begin{split}
 & \left\{\theta_{0,0},\theta_{0,1},\theta_{0,2},\theta_{0,3},\theta_{0,4}\right\} = 
  \left\{
 1, 1 + \frac{D}{2 \gamma }, 1 + \frac{D}{\gamma}, 1 + \frac{3 D}{2\gamma}, 
  1 +  \frac{2 D}{\gamma}
\right\}, \\
  & \left\{\theta_{1,1},\theta_{1,2},\theta_{1,3},\theta_{1,4}\right\} = 
   \left\{
   \frac{1}{\gamma},  \frac{3}{\gamma}, \frac{6}{\gamma}, \frac{10}{\gamma}
  \right\}, \\
       &  \left\{\theta_{2,3},\theta_{2,4}\right\} = 
       \left\{
       \frac{3}{\gamma ^2},  \frac{15}{\gamma ^2}
       \right\}.
 \end{split}
 \end{align}
 Therefore, in the classical  limit, we have
 \begin{align}\label{27c}
     \frac{N_3}{D_3} =  2, \qquad  \frac{N_{31}}{D_3}=\frac{10}{2+D}, \qquad C_5 = 1,
 \qquad  \frac{N^\Pi}{D_4} = - 1, \qquad  \frac{N^\Delta}{D_4} = -  \frac{2}{D},
 \end{align}
 and we find from \eqref{I8nn} 
  \begin{align}\label{classi}
      \begin{split}
      {\bar{\chi}}_\text{class} = \frac{D+2}{2},\qquad
           {\bar{\nu}} _\text{class} =  \frac{2(D-3)}{3D}     
             , \qquad
           {\bar{\mu} }_\text{class} =1,
           \end{split}
  \end{align}
which are in perfect agreement with the phenomenological coefficients of the classical RET theory \cite{RS}.
 
\subsection{Phenomenological coefficients in ET$_{14}$ and ET$_{6}$}

By conducting the Maxwellian iteration to ET$_{14}$  as a principal subsystems of ET$_{15}$, we may expect that the different bulk viscosity appears. This is because $\Delta$ is related to $\Pi$  by \eqref{subd} and it affects the balance laws corresponding to $\Pi$ in ET$_{14}$. In fact, from \eqref{Sub2} and \eqref{Sub3}, we can obtain the closed field equations for $\Pi$, and then, through the Maxwellian iteration as it has been done in \cite{Car2}, we obtain the bulk viscosity for ET$_{14}$ as follows:
\begin{align}\label{nu14}
 \bar{\nu}_{14} = \frac{\frac{1}{\omega'}\left(\theta_{0,2}' + \frac{1}{3}\theta_{1,2}'\right) - \frac{8}{9}\gamma \theta_{1,2}}{\left(-1 + \frac{N^\Delta}{D_4}\right)\frac{N_a}{D_a} + \frac{N^\Pi}{D_4}}.
\end{align}
We remark that the heat conductivity and the shear viscosity is the same between  ET$_{15}$ and ET$_{14}$.



Similarly, from \eqref{field6bis}$_4$, we obtain the bulk viscosity estimated by ET$_6$ as follows:
\begin{align} \label{nu6}
	& \bar{\nu}_6 = - \frac{\theta_{0,2}' - \theta_{1,2}'}{\omega' A_1}.
\end{align}

It should be noted that, in the classical case studied in \cite{x}, the bulk viscosities of ET$_{15}$, ET$_{14}$ and ET$_{6}$ are the same. In fact, in the classical limit, $\bar{\nu}_{14}$ and $\bar{\nu}_6$ coincides with $\bar{\nu}_\text{class}$. However, due to the mathematical structure of the relativity, i.e., the scalar fields $\Pi$ and $\Delta$ appear together in the triple tensor, the method of the principal subsystem dictates the difference of the subsystems.

\subsection{Heat conductivity, Bulk viscosity and Shear viscosity in diatomic gases}
Inserting \eqref{omegaD}, after cumbersome calculations (easy with Mathematica), we can obtain the phenomenological coefficients in the diatomic case:

{\small 
\begin{align*}
&\bar{\chi}=-\frac{\gamma  \Big(\gamma^2+2 \gamma  G-8\Big)
   \left\{\gamma ^4
   \Big(G^2-1\Big)+2 \gamma
   ^2 \Big(G^2+2\Big)-5
   \gamma ^3 G-16 \gamma 
   G+32\right\}^2}
{(\gamma  G-4)^3 \left\{\gamma \left[-\gamma ^5+5 \gamma
   ^3+48 \gamma +\Big(\gamma
   ^4-6 \gamma ^2-12\Big)
   \gamma  G^2+\Big(-5 \gamma
   ^4+12 \gamma ^2+96\Big)
   G\right]-192\right\}}, \\
&   \bar{\mu} = \frac{\Big(\gamma ^2+2 \gamma G-8\Big)^2}{(\gamma  G-4)
      \left\{4 \Big(\gamma
      ^2-8\Big)+\gamma 
      \Big(\gamma ^2+8\Big)
      G\right\}},\\
      &\bar{\nu} = \frac{g_1}{3 (\gamma  G-4) g_2}, 
      \end{align*}
      with
      \begin{align*}
      &g_1= 4
               \gamma ^{15} G
               \Big(G^2-1\Big)^2+81920
               \gamma ^3 G \Big(7
               G^2+20\Big)-196608 \gamma^2 
               \Big(7 G^2+4\Big)+1024
                              \gamma ^5 G \Big(21 G^4+660
                              G^2-392\Big) -\\
               & \hspace{1cm} 4096 \gamma
               ^4 \Big(35 G^4+348
               G^2-56\Big)+4 \gamma ^{14}
               \Big(G^6-17 G^4+21
               G^2-5\Big)+\gamma ^{13} G
               \Big(7 G^6-86 G^4+435
               G^2-256\Big)+\\
           & \hspace{1cm}      
               4 \gamma
               ^{12} \Big(-40 G^6+193
               G^4-331 G^2+48\Big)+4
               \gamma ^{11} G \Big(-14
               G^6+422 G^4-943
               G^2+500\Big)+16 \gamma ^{10}
               \Big(77 G^6-660 G^4+677
               G^2-84\Big)+\\
                & \hspace{1cm} 
               16 \gamma ^9 G
               \Big(7 G^6-714 G^4+2560
               G^2-1108\Big)-64 \gamma ^8
               \Big(45 G^6-910 G^4+1472
               G^2-204\Big)+ \\
                & \hspace{1cm} 64 \gamma ^7
               G \Big(G^6+492 G^4-2800
               G^2+1760\Big)-256 \gamma
               ^6 \Big(7 G^6+740 G^4-1344
               G^2+192\Big)+1835008 \gamma G - 1048576,  \\
                & g_2 = \gamma ^4
                  \Big(G^2-1\Big)+\gamma ^2
                  \Big(G^2+4\Big)-5 \gamma
                  ^3 G-8 \gamma  G+16\Big)
                  \Big(\gamma  \Big(2 \gamma
                  ^9 G^2 \Big(G^2-1\Big)+5
                                    \gamma ^8 G \Big(1-3
                                    G^2\Big)+\\
                   &\hspace{1cm} 40 \gamma ^6 G
       \Big(6-5 G^2\Big)+64
                  \gamma ^4 G \Big(11
                  G^2-25\Big)+512 \gamma ^2
                  G \Big(G^2+14\Big)-1024
                  \gamma  \Big(3
                  G^2+5\Big)+\\
                 &\hspace{1cm}  \gamma ^7
                  \Big(19 G^4-17
                  G^2+28\Big)-4 \gamma ^5
                  \Big(13 G^4-198
                  G^2+60\Big)-32 \gamma ^3
                  \Big(G^4+108
                  G^2-52\Big)+8192
                  G\Big)-8192.
\end{align*}
}
\normalsize

\begin{figure}[h]
\includegraphics[width=0.48\linewidth]{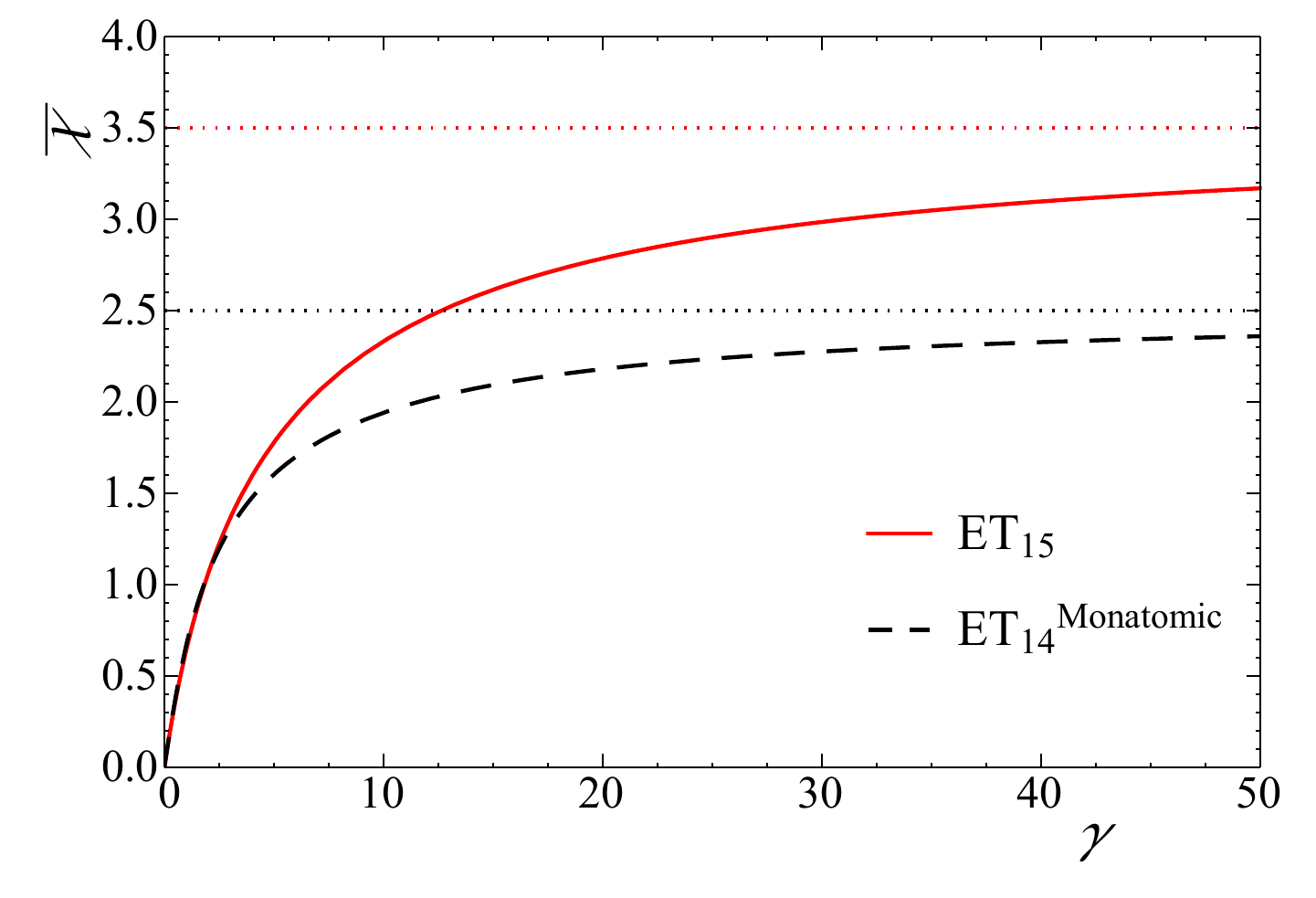} \hspace{0.5cm} \includegraphics[width=0.48\linewidth]{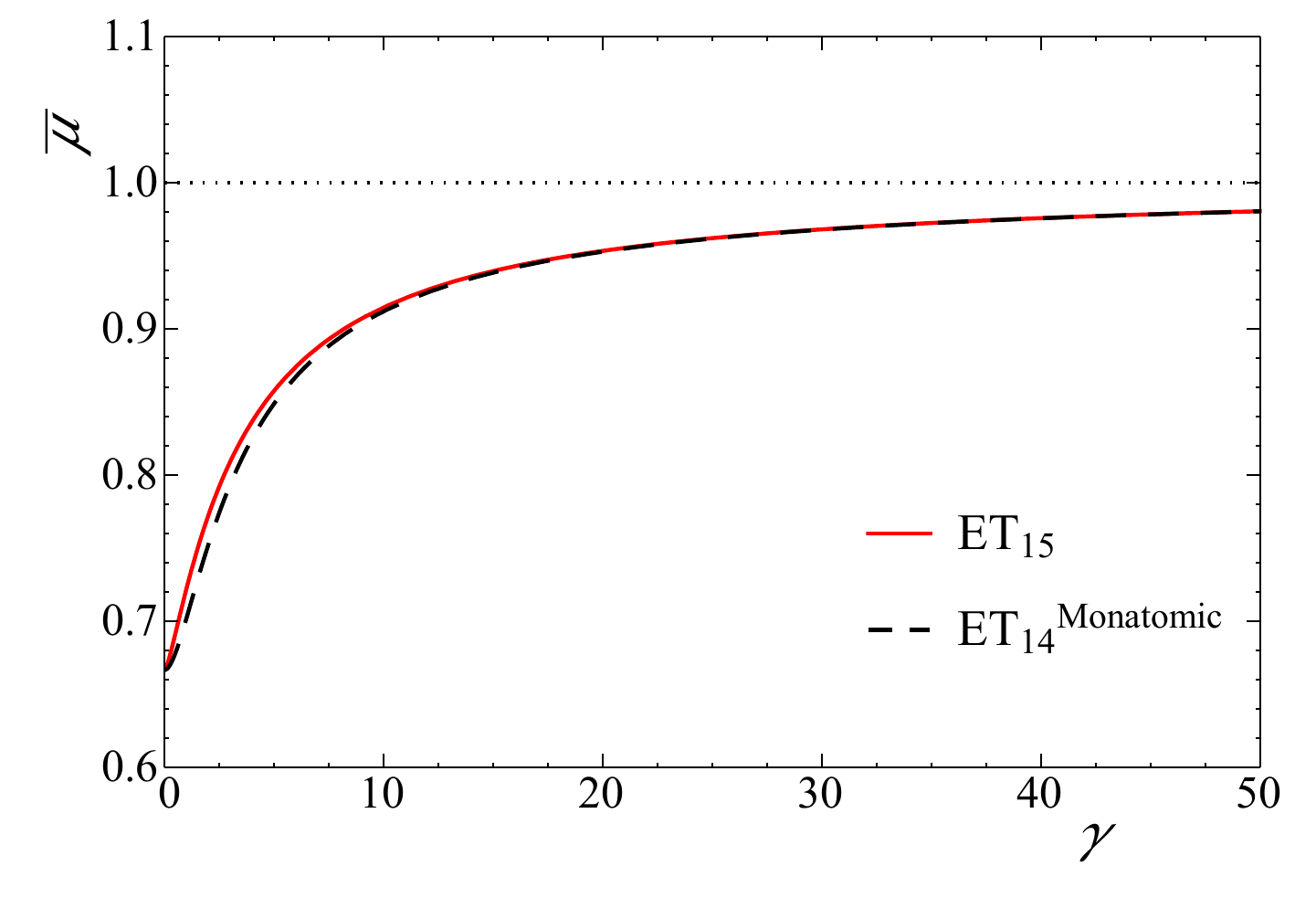}
\caption{Dependence of $\bar{\chi}$ (left) and $\bar{\mu}$ (right) for diatomic (red solid line) and monatomic (black dashed line) gases on $\gamma$. The dotted line indicates the corresponding value in the classical limit. In the ultra-relativistic limit ($\gamma \to 0$), $\bar{\chi}_\text{ultra}=0, \bar{\mu}_\text{ultra}=2/3$ both for monatomic and diatomic gases. In the classical limit ($\gamma \to \infty$), $\bar{\chi}_\text{class}=2.5, \bar{\mu}_\text{class}=1$ for monatomic gas, and $\bar{\chi}_\text{class}=3.5, \bar{\mu}_\text{class}=1$ for diatomic gas.}
\label{fig:chidiamono}
\end{figure}

\begin{figure}[h]
\includegraphics[width=0.48\linewidth]{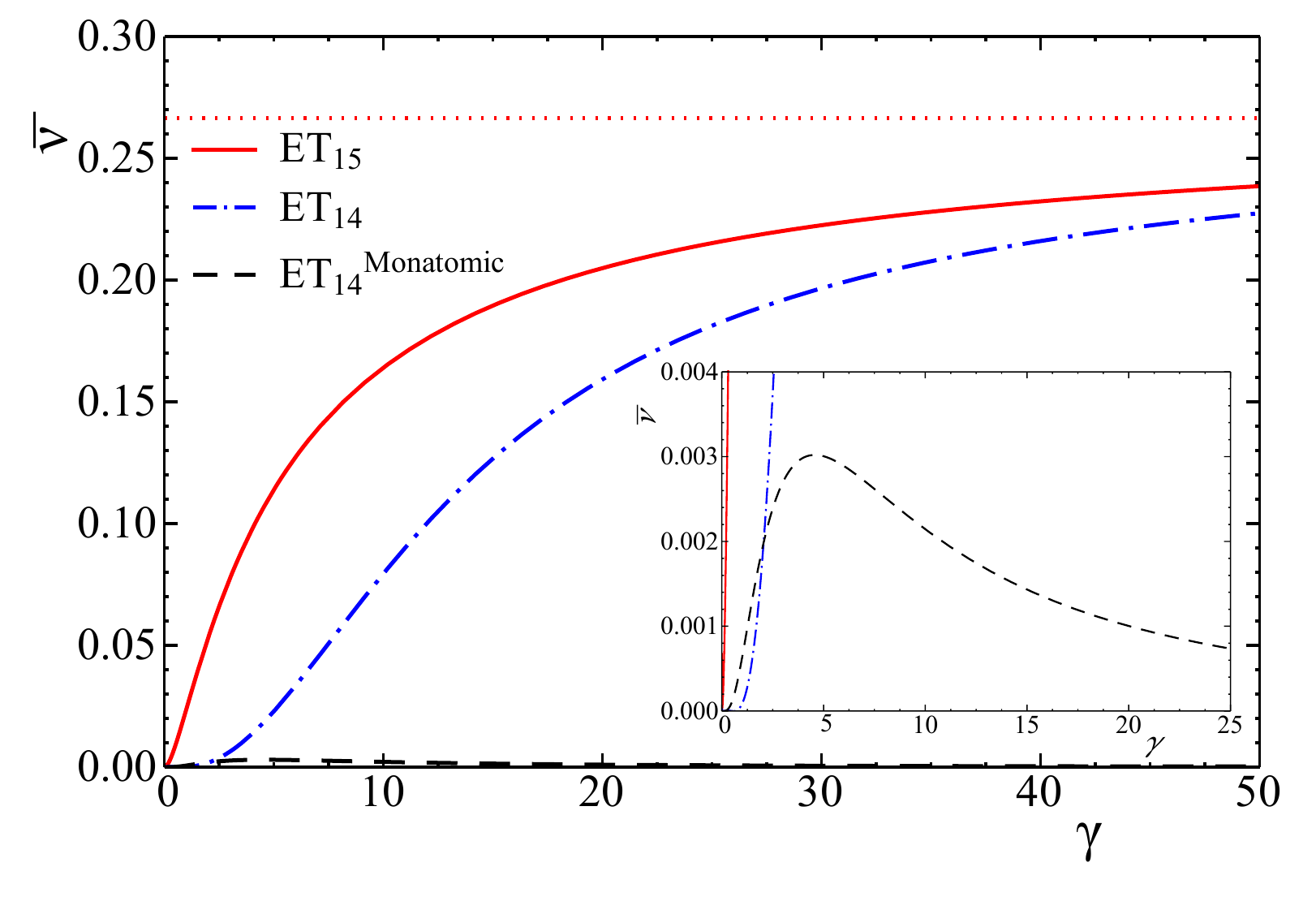} 
\caption{Dependence of $\bar{\nu}$ for  diatomic (red solid line)  monatomic (black dashed line)  gases on $\gamma$. The prediction by $ET_{14}$ as a principal subsystem of ET$_{15}$ is also shown with the dotted line. In the ultra-relativistic limit ($\gamma \to 0$), $\bar{\nu}_\text{ultra}=0$ both for monatomic and diatomic gases. In the classical limit ($\gamma \to \infty$), $\bar{\nu}_\text{class}=0$ for monatomic gas, and $\bar{\nu}_\text{class}=4/15$ for diatomic gas.}
\label{fig:nubarr}
\end{figure}

Let us compare the phenomenological coefficients with the ones  for the  monatomic case obtained in \cite{Car2}. In Fig.\ref{fig:chidiamono}, we plot the dependence of the dimensionless heat conductivity and shear viscosity on $\gamma$ for both diatomic and monatomic cases. Concerning $\nu$, we also plot the dimensionless bulk viscosity of ET$_{14}$ derived in \eqref{nu14} in the Fig.\ref{fig:nubarr}. We observe that in the ultra-relativistic limit and in classical limit, the figures are in perfect agreement with the limits \eqref{I8ultran} and \eqref{classi} (for $D=3,5$). We remark, as it is evidently shown in Fig.\ref{fig:nubarr}, how small is the bulk viscosity in monatomic gas with respect to the one of diatomic case. 

It is also remarkable that the value of the bulk viscosity of ET$_6$ given by \eqref{nu6} is quite near to the one of ET$_{15}$. For this reason, we omit the plot of $\bar{\nu}^{(6)}$ in the figure. This indicates that ET$_6$ captures the effect of the dynamic pressure in consistent with ET$_{15}$.

\section{Classic limit of the relativistic theory}\label{sec:classical}
We want to perform now the classical limit $\gamma \rightarrow \infty $ of the closed relativistic system \eqref{derivmat}. For this purpose we recall the limits of the coefficients given in \eqref{thetaclass} and \eqref{27c}. Moreover, 
taking into account the decomposition $U^\alpha \, \equiv \, \Big(\Gamma \, c \, , \, v^i\Big)$, where $\Gamma$ is the Lorentz factor, we have $\partial_\alpha U^\alpha = \, \frac{1}{c} \, \partial_t \Big(\Gamma \, c\Big) + \, \partial_k \, \Big(\Gamma \, v^k\Big)$ whose limit is $\partial_i v^i$ because $\partial_t \, \Gamma = - \, \Gamma^3 \frac{v_i}{c^2} \, \partial_t v^i$ has zero limit, and the similar evaluation applies to $\partial_k \, \Gamma$. Then, 
 \begin{align*}
    \frac{1}{c^2} \, U^\mu \partial_\mu U^0 =  \frac{1}{c^2} \, \Gamma \, c \, \frac{1}{c} \, \partial_t \, \Big( \Gamma \, c \Big) \, + \,
    \frac{1}{c^2} \, \Gamma \, v^k \, \partial_k \, \Big( \Gamma \, c \Big) \quad \mbox{has 0 limit} \, , \\
    \frac{1}{c^2} \, U^\mu \partial_\mu U^i =  \frac{1}{c^2} \, \Gamma \, c \, \frac{1}{c} \, \partial_t \, \Big( \Gamma \, v^i \Big) \, + \,
    \frac{1}{c^2} \, \Gamma \, v^k \, \partial_k \, \Big( \Gamma \, v^i \Big) \quad \mbox{has 0 limit} \, .
 \end{align*}
Concerning the projection operator in the limit, it is necessary to remind that, with our choice of the metric, $v_j= - \, v^j$, then
\begin{align*}
    h^{\beta \alpha} =  &- \, g^{\beta \alpha} + \frac{U^\beta U_\alpha}{c^2} \quad \rightarrow \quad h^{i j} = - \, g^{i j} + \Gamma^2 \, \frac{v^i v^j}{c^2} \quad \rightarrow \quad  \lim_{c \, \rightarrow \, + \infty} \, h^{i j} = - \, g^{i j} =   \delta^{i j} , \\
  &   \lim_{c \, \rightarrow \, + \infty} \, h^{i}_j = - \, g^{i}_j =  - \, \mbox{diag} \, (1, \, 1, \, 1) \, . \nonumber
\end{align*}
While from
\begin{align*}
    \begin{split}
        & 0 = U_\alpha h^{i \alpha} = \Gamma \, c \, h^{i 0} +  \Gamma \, v_k \, h^{i k} \quad \rightarrow \quad  h^{i 0} = -  \frac{v_k}{c} \, h^{i k} \, , \\
        & \, 0 = U_\alpha h^{0 \alpha} = \Gamma \, c \, h^{0 0} +  \Gamma \, v_k \, h^{0 k} \quad
        \rightarrow \quad  h^{0 0} = -  \frac{v_k}{c} \, h^{0 k} =  -  \frac{v_a v_b}{c^2} \, h^{a b}\, .
    \end{split}
\end{align*}
The last two relations hold also without taking the non relativistic limit. As a consequence, we have that $ \lim_{c \, \rightarrow \, + \infty} \, h^{i 0} = 0$ and  $\lim_{c \, \rightarrow \, + \infty} \, h^{00} = 0$.

The relativistic material derivative \eqref{matde}  of a function $f$ converges to the classical material derivative where we continue to indicate it with a dot. Then, the system \eqref{derivmat} becomes in the classical limit:
\begin{align}
\begin{split}
 &\dot{\rho} + \rho \frac{\partial v_l}{\partial x_l} = 0,\\
 &\rho \dot{v}_i + \frac{\partial p}{\partial x_i} + \frac{\partial \Pi}{\partial x_i} - \frac{\partial \sigma_{\langle ik\rangle}}{\partial x_k} =0,\\
 &\dot{T} + \frac{2T}{Dp} \left\{(p+\Pi)\frac{\partial v_l}{\partial x_l} - \sigma_{\langle ik\rangle}\frac{\partial v_k}{\partial x_i}+\frac{\partial q_l}{\partial x_l}\right\}
 =0,\\ 
 &\dot{\Pi} + \frac{2}{3}\frac{D -3}{D}p \frac{\partial v_l}{\partial x_l} + \frac{5D-6}{3D}\Pi \frac{\partial v_l}{\partial x_l} - \frac{2}{3}\frac{D-3}{D}\sigma_{\langle lk\rangle}\frac{\partial v_l}{\partial x_k}  + \frac{4(D-3)}{3D(D+2)}\frac{\partial q_l}{\partial x_l} = - \frac{1}{\tau  }\, \Pi,\\
 &\dot{\sigma}_{\langle ij\rangle} + \sigma_{\langle ij\rangle}\frac{\partial v_l}{\partial x_l} + 2\sigma_{\langle l\langle i\rangle}\frac{\partial v_{j\rangle}}{\partial x_l} - 2 (p+\Pi)\frac{\partial v_{\langle j}}{\partial x_{i\rangle}}   - \frac{4}{D+2} \frac{\partial q_{\langle i}}{\partial x_{j\rangle}} = -  \frac{1}{\tau  } \sigma_{\langle ij\rangle},\\
 & \dot{q}_i + \frac{D+4}{D+2}q_i \frac{\partial v_l}{\partial x_l} + \frac{D+4}{D+2} q_l \frac{\partial v_i}{\partial x_l} + \frac{2}{D+2}q_l \frac{\partial v_l}{\partial x_i} \\
& \qquad + \frac{D+2}{2}\frac{p}{\rho T}\left\{\left(p+\Pi \right)\delta_{il} - \sigma_{\langle il\rangle}\right\}\frac{\partial T}{\partial x_l}  - \frac{p}{\rho^2} \left(\Pi \delta_{il} - \sigma_{\langle il\rangle}\right)\frac{\partial \rho}{\partial x_l}  \\
&\qquad + \frac{1}{\rho}\left\{(p-\Pi)\delta_{il}+ \sigma_{\langle il\rangle} \right\}\left(\frac{\partial \Pi}{\partial x_l}  - \frac{\partial \sigma_{\langle rl\rangle}}{\partial x_r}\right) +\frac{1}{2D}\frac{\partial \Delta}{\partial x_i} = -  \frac{1}{\tau  }\, q_i,\\
 & \dot{\Delta} + \left(\frac{D+4}{D}\Delta + 8\frac{p}{\rho}\Pi \right)\frac{\partial v_l}{\partial x_l}  - 8 \frac{p}{\rho}\sigma_{\langle ik\rangle}\frac{\partial v_i}{\partial x_k} - \frac{8}{\rho}q_i \frac{\partial p}{\partial x_i} \\
 &\qquad + 4(D+4)\frac{p}{\rho T}q_l \frac{\partial T}{\partial x_l}  
 + \frac{8p}{\rho} \frac{\partial q_l}{\partial x_l} - \frac{8}{\rho}q_i \frac{\partial \Pi}{\partial x_i} + \frac{8}{\rho} q_i \frac{\partial \sigma_{\langle il\rangle}}{\partial x_l}  =   - \frac{1}{\tau  }\Delta,
\end{split}
\label{fieldeqsMpoly}
\end{align}
where $\sigma_{\langle ij\rangle} = -t_{\langle ij\rangle}$. The system \eqref{fieldeqsMpoly} coincides perfectly with the one obtained recently in \cite{x}.

We remark that, as it has been studied in \cite{x}, for classical polytropic gases, ET$_{14}$ is derived as a principal subsystem of ET$_{15}$  by  setting $\Delta=0$. Moreover, ET$_6$ is derived from ET$_{14}$  as a principal subsystem of ET$_{14}$  by  setting $\sigma_{\langle ij\rangle}=0$ and $q_i=0$.  This fact corresponds that, in the classical limit, both of $\Delta^{(14)}$ defined in \eqref{subd} and $\Delta^{(6)}$ defined in \eqref{subd6} become zero.

\bigskip
\noindent

\acknowledgments{
 The work has been partially supported by JSPS KAKENHI Grant Numbers JP18K13471 (TA),   by the Italian MIUR through the PRIN2017
 project "Multiscale phenomena in Continuum Mechanics:
 singular limits, off-equilibrium and transitions" Project Number:
 2017YBKNCE (SP) and GNFM/INdAM (MCC, SP and TR).
}

\conflictsofinterest{The authors declare no conflict of interest.}

\appendixtitles{no} 
\appendix

\section{A. Entropy-entropy flux density} \label{app:entropy}
In order to evaluate the entropy density from eq. \eqref{entropy}, we need the expression of $f\, \ln f$ up to the second order with respect to the nonequilibrium variables.\\
The expansion of the distribution function around an equilibrium state is 
\begin{align*}
	\begin{split}
		& f=f_E \, \, e^{\frac{-1}{k_B} \,\tilde{\chi}} =  f_E \, \left[1 \, - \, \frac{1}{k_B} \,  \tilde{\chi}  \, + \, \frac{1}{2 \, k_B^2} \, \left(\tilde{\chi}\right)^2  \, +  \, \left(\tilde{\chi}\right)^3 ( \cdots) \right] \, , \\
		& \mbox{with} \quad \tilde{\chi} = \left(\tilde{\chi}\right)^{(1)} + \left(\tilde{\chi}\right)^{(2)} + \left(\tilde{\chi}\right)^{(3)} \, ( \cdots) \, , 
	\end{split}
\end{align*}
defined in \eqref{fgenE}$_2$ and the notation $\eta^{(i)}$ represents the homogeneous part of the generic quantity $\eta$ at the order $i$ with respect to the nonequilibrium variables. With this notation, the quantities $\left( \lambda-\lambda^E\right)^{(1)}$, $\left(\lambda_{\beta}-\lambda_{\beta}^E\right)^{(1)}$, $\left(\lambda_{\beta \gamma }\right)^{(1)}$ are those of eq. \eqref{RT}.\\
By composing the above expressions, we see that the distribution function up to the second order is
\begin{align*}
	f=  f_E \, \left\{ 1 \, - \, \frac{1}{k_B} \, \left[\left(\tilde{\chi}\right)^{(1)}  + \left(\tilde{\chi}\right)^{(2)} \right] + \, \frac{1}{2 \, k_B^2} \, \left[\left(\tilde{\chi}\right)^{(1)} \right]^2 \right\} \, .
\end{align*}
and 
\begin{align*}
	\begin{split} 
		 f \, \ln f =& f \, \left(-1 - \, \frac{1}{k_B} \, \tilde{\chi}\right)=  f_E \, \left\{ 1 \, - \, \frac{1}{k_B} \, \left[\left(\tilde{\chi}\right)^{(1)}  + \left(\tilde{\chi}\right)^{(2)} \right] + \, \frac{1}{2 \, k_B^2} \, \left[\left(\tilde{\chi}\right)^{(1)} \right]^2 + \cdots \right\} \cdot \\
		& \hspace{3cm}  \left\{ -1 - \, \frac{1}{k_B} \, \chi_E \, - \, \frac{1}{k_B} \, \left[\left(\tilde{\chi}\right)^{(1)}  + \left(\tilde{\chi}\right)^{(2)} + \cdots \right] \right\} \\ 
		=& f_E \, \ln \, f_E \, + \, \frac{1}{k_B^2} \, f_E \, \chi_E \, \left(\tilde{\chi}\right)^{(1)} \, + \, \frac{1}{k_B^2} \, f_E \, \chi_E \, \left\{\left(\tilde{\chi}\right)^{(2)} \, - \, \frac{1}{2 \, k_B} \, \left[\left(\tilde{\chi}\right)^{(1)} \right]^2 \right\} + \\
		& + \, \frac{1}{2 \, k_B^2} \, f_E \, \left[\left(\tilde{\chi}\right)^{(1)} \right]^2 \, .
	\end{split} 
\end{align*}
It follows that
\begin{align*}
	h^\alpha  = - k_B \, c \int_{\R^3} \int_{0}^{+ \infty} p^\alpha f \, \ln \, f \, \varphi ( \mathcal{I} ) \, d \, \mathcal{I} \, d \, \boldsymbol{P} =  h^\alpha_E \, + \,  h^\alpha_{(1)} \, + \,  h^\alpha_{(2)}  \, ,
\end{align*}
where 
\begin{align*}
	\begin{split} 
		& h^\alpha_{(1)}  = -\frac{c}{k_B}  \int_{\R^3} \int_{0}^{+ \infty} p^\alpha f_E \, \chi_E \, \left(\tilde{\chi}\right)^{(1)} \, \varphi ( \mathcal{I} ) \, d \, \mathcal{I} \, d \, \boldsymbol{P} =   \\
		&  = - \frac{c}{k_B}  \int_{\R^3} \int_{0}^{+ \infty} p^\alpha f_E \, \left[ m \, \lambda_E + \left(1 + \frac{\mathcal{I}}{m \, c^2}\right) \, \frac{U_\mu}{T} \, p^\mu\right] \, \left(\tilde{\chi}\right)^{(1)} \, \varphi ( \mathcal{I} ) \, d \, \mathcal{I} \, d \, \boldsymbol{P},
	\end{split} 
\end{align*}
\begin{align*}
	\begin{split} 
		& h^\alpha_{(2)}  = - \frac{c}{k_B}  \int_{\R^3} \int_{0}^{+ \infty} p^\alpha f_E \, \chi_E \, \left\{\left(\tilde{\chi}\right)^{(2)} \, - \, \frac{1}{2 \, k_B} \, \left[\left(\tilde{\chi}\right)^{(1)} \right]^2 \right\}  \, \varphi ( \mathcal{I} ) \, d \, \mathcal{I} \, d \, \boldsymbol{P} \,  -\\
		& \quad \quad  \,  \frac{c}{2 \, k_B}  \int_{\R^3} \int_{0}^{+ \infty} p^\alpha f_E \, \left[\left(\tilde{\chi}\right)^{(1)} \right]^2  \, \varphi ( \mathcal{I} ) \, d \, \mathcal{I} \, d \, \boldsymbol{P}.
	\end{split} 
\end{align*}
Moreover, we have that the moments appearing in system \eqref{Annalis} up to the second order as follows:
\begin{align*}
	\begin{split}
		& \underline{V^\alpha =  V^\alpha_E} - \, \frac{m \, c}{k_B} \int_{\R^3} \int_{0}^{+ \infty} p^\alpha f_E \left\{ \underline{\left(\tilde{\chi}\right)^{(1)}}  + \left(\tilde{\chi}\right)^{(2)} - \, \frac{1}{2 \, k_B} \, \left[\left(\tilde{\chi}\right)^{(1)} \right]^2  \right\} \varphi ( \mathcal{I} ) \, d \, \mathcal{I} \, d \, \boldsymbol{P} \, , \\
		& \underline{T^{\alpha \beta} =  T^{\alpha \beta}_E} - \, \frac{c}{k_B} \int_{\R^3} \int_{0}^{+ \infty} p^\alpha p^\beta \left(1 + \frac{\mathcal{I}}{m \, c^2}\right) f_E \left\{ \underline{\left(\tilde{\chi}\right)^{(1)}}  + \left(\tilde{\chi}\right)^{(2)} - \, \frac{1}{2 \, k_B} \, \left[\left(\tilde{\chi}\right)^{(1)} \right]^2  \right\} \varphi ( \mathcal{I} ) \, d \, \mathcal{I} \, d \, \boldsymbol{P}\, , \\
		&\underline{ \frac{U_ \alpha U_\beta U_\gamma}{c^4} \, A^{\alpha \beta \gamma} =  \frac{U_ \alpha U_\beta U_\gamma}{c^4} \, A^{\alpha \beta \gamma}_E} - \\
		& \frac{U_ \alpha U_\beta U_\gamma}{c^4} \,  \frac{c}{m \, k_B} \int_{\R^3} \int_{0}^{+ \infty} p^\alpha p^\beta p^\gamma \left(1 + \frac{\mathcal{I}}{m \, c^2}\right)^2 f_E \left\{ \underline{\left(\tilde{\chi}\right)^{(1)} } + \left(\tilde{\chi}\right)^{(2)} - \, \frac{1}{2 \, k_B} \, \left[\left(\tilde{\chi}\right)^{(1)} \right]^2  \right\} \varphi ( \mathcal{I} ) \, d \, \mathcal{I} \, d \, \boldsymbol{P}\, .
	\end{split} 
\end{align*}
The underlined terms gives 0 for eqs. \eqref{RT} and there remain
\begin{align*}
	\begin{split}
		&  - \, \frac{m \, c}{k_B} \int_{\R^3} \int_{0}^{+ \infty} p^\alpha f_E \left\{   \left(\tilde{\chi}\right)^{(2)} - \, \frac{1}{2 \, k_B} \, \left[\left(\tilde{\chi}\right)^{(1)}\right]^2  \right\} \varphi ( \mathcal{I} ) \, d \, \mathcal{I} \, d \, \boldsymbol{P}=0 \, , \\
		&   - \, \frac{c}{k_B} \int_{\R^3} \int_{0}^{+ \infty} p^\alpha p^\beta \left(1 + \frac{\mathcal{I}}{m \, c^2}\right) f_E \left\{  \left(\tilde{\chi}\right)^{(2)} - \, \frac{1}{2 \, k_B} \, \left[\left(\tilde{\chi}\right)^{(1)} \right]^2  \right\} \varphi ( \mathcal{I} ) \, d \, \mathcal{I} \, d \, \boldsymbol{P} =0 \, , \\
		& \frac{U_ \alpha U_\beta U_\gamma}{c^4} \,  \frac{c}{m \, k_B} \int_{\R^3} \int_{0}^{+ \infty} p^\alpha p^\beta p^\gamma \left(1 + \frac{\mathcal{I}}{m \, c^2}\right)^2 f_E \left\{  \left(\tilde{\chi}\right)^{(2)} - \, \frac{1}{2 \, k_B} \, \left[\left(\tilde{\chi}\right)^{(1)} \right]^2  \right\} \varphi ( \mathcal{I} ) \, d \, \mathcal{I} \, d \, \boldsymbol{P} = 0\, .
	\end{split} 
\end{align*}
The first two of them allows prove eq. \eqref{entropyd2}$_1$ and to write 
\begin{align*}
	\begin{split} 
		& h^\alpha_{(2)} =  -\,  \frac{c}{2 \, k_B}  \int_{\R^3} \int_{0}^{+ \infty} p^\alpha f_E \, \left[\left(\tilde{\chi}\right)^{(1)} \right]^2  \, \varphi ( \mathcal{I} ) \, d \, \mathcal{I} \, d \, \boldsymbol{P} \, .
	\end{split} 
\end{align*}
It is sufficient to substitute the expression of $\tilde{\chi}$ to obtain eq. \eqref{entropyd2}$_2$

\section{B. Continuity of the ultra relativistic limit for $a=2$}\label{app:cont}
From \eqref{10}$_2$, and by using the recurrence relations \eqref{Rbis} and \eqref{R}, we have
\begin{align*}
    \frac{\gamma e}{n \, m \, c^2}=
    \frac{\gamma \, \int_0^{+\infty} J_{2,2}^* \, \Big( 1 + \frac{\mathcal{I}}{m c^2} \Big) \, \phi(\mathcal{I})  \, d \, \mathcal{I}}{\int_0^{+\infty} J_{2,1}^* \,  \phi(\mathcal{I})  \, d \, \mathcal{I}}= 3 +
    \frac{\int_0^{+\infty} J_{0,1}^* \, \phi(\mathcal{I})  \, d \, \mathcal{I}}{\int_0^{+\infty} J_{2,1}^* \,  \phi(\mathcal{I})  \, d \, \mathcal{I}}.
\end{align*}
By introducing the Ruggeri's numbers $R_k$ and using eq. (32)$_{1}$ of \cite{6} we have
\begin{align*}
    \frac{\gamma e}{n \, m \, c^2}= 3 + \frac{3}{- \, \ln \, \gamma} \, R_{-4} = 3 - \frac{1}{ \ln \gamma}
\end{align*}
or
\begin{align}\label{28}
    e =  \frac{n \, m \, c^2}{\gamma} \Big( 3 - \, \frac{1}{\ln \, \gamma} \Big)\, .
\end{align}
So we have to calculate $D_4$, $N^\Pi$ and $N^\Delta$ with \eqref{28} instead of \eqref{25}.

In particular, for $D_4$ we can add to its fourth line the second one pre-multiplied by $- \, \frac{1}{3} $ so that it becomes
\begin{align*}
    \frac{1}{\gamma\, \ln \gamma} \left( \frac{1}{3},
    \frac{4}{3\,\gamma},  \frac{20}{3\,\gamma^2 }, \frac{4}{3} \,   \frac{c^2}{\gamma^2 } \right) \, .
\end{align*}
It follows that, after cumbersome calculations that we do not report here for brevity, we have
\begin{align*}
    \lim_{\gamma \, \rightarrow \, 0} \gamma^{9} \, \ln \, \gamma \,  D_4 =  \left|
    \begin{matrix} 1  & 3  & 12  & 4  \\
        &&&& \\
        3 & 12  & 60 & 20\\
        &&&& \\
        12   & 60  & 360  &  120   \\
        &&&& \\
        \frac{1}{3}   & \frac{4}{3}  & \frac{20}{3} & \frac{4}{3}
    \end{matrix}
    \right| = - 64 \, .
\end{align*}
Similarly, for $N^\Pi$ we can add to its fourth line the third one multiplied by $- \, \frac{1}{3} $ so that it becomes
\begin{align*}
    \frac{1}{\gamma^2 \, \ln \gamma} \left( \frac{4}{3},
    \frac{20}{3\, \gamma},  \frac{40}{\gamma^2 } ,  \frac{8\, c^2}{\gamma^2 } \right) \, .
\end{align*}
It follows that
\begin{align*}
    \lim_{\gamma \, \rightarrow \, 0} \gamma^{10} \, \ln \, \gamma \,  N^\Pi = - \, \left|
    \begin{matrix} 1  & 3  & 12  & 4  \\
        &&&& \\
        3 & 12  & 60 & 20\\
        &&&& \\
        12   & 60  & 360  &  120   \\
        &&&& \\
        \frac{4}{3}   & \frac{20}{3}  & 40 & 8
    \end{matrix}
    \right| = 384 \, .
\end{align*}
Finally, for $N^\Delta$ we can add to its third line the second one multiplied by $- \, \frac{1}{3} $ so that its third line becomes
\begin{align*}
    \frac{1}{\gamma \, \ln  \gamma}   \left( \frac{1}{3},
    \frac{4}{3 \, \gamma},  \frac{20}{3\, \gamma^2}, \frac{4}{3}   \frac{c^2}{\gamma^2} \right) \, .
\end{align*}
It follows that
\begin{align*}
    \lim_{\gamma \, \rightarrow \, 0} \gamma^{9} \, \ln \, \gamma \, N^\Delta =  \left|
    \begin{matrix} 1  & 3  & 12  & 4  \\
        &&&& \\
        3 & 12  & 60 & 20\\
        &&&& \\
        \frac{1}{3}   & \frac{4}{3}  & \frac{20}{3} &   \frac{4}{3} \\
        &&&& \\
        4 & 20 & 120 & 40
    \end{matrix}
    \right| = \frac{64}{3} \, .
\end{align*}
By joining all these results we obtain
\begin{align*}
    \lim_{\gamma \, \rightarrow \, 0} \frac{\gamma \, N^\Pi}{D_4} = - 6 \, ,
    \quad \lim_{\gamma \, \rightarrow \, 0} \frac{N^\Delta}{D_4} = - \, \frac{1}{3}
    \, ,
\end{align*}
which confirms \eqref{26} also for $a=2$.

\end{document}